# Intercity mobility reveals the hyperbolic geometry of city systems

Zhaoya Gong[1,2], Bin Liu[1,2,†], Chenglong Wang[1,2,†], Pengjun Zhao[1,2,*], Xiang Li[1,2], Kaixiang Zhang[1,2], Changcheng Kan[1,2], Xingjian Liu[3]

[1] School of Urban Planning and Design, Peking University Shenzhen Graduate School, Shenzhen, China

[2] Key Laboratory of Earth Surface System and Human-Earth Relations of Ministry of Natural Resources of China, Peking University Shenzhen Graduate School, Shenzhen, China

[3] Department of Urban Planning and Design, The University of Hong Kong, Hong Kong, China

† Contributed equally

* Corresponding author: Pengjun Zhao. Email: pengjun.zhao@pku.edu.cn

**Abstract**

The hierarchy and proximity are key dimensions of urban relational processes, but their interplay in shaping intercity interactions and the underlying structures of city systems remain unclear. We develop a novel geometric model of city systems embedding intercity mobility into a latent hyperbolic geometry, which unravels the measures of hierarchy and proximity accounting for their interplay. It is successfully validated against 12 different nationwide intercity mobility datasets. We find a bottom-up emergence of city hierarchies, along which the variations of city-hinterland relations are non-stationary in terms of their nesting and range properties. Such non-stationarity originates from trade-offs between city hierarchy and hinterland range in determining the formation of city-hinterland structures. Hierarchy- and proximity-dominated urban processes can be elucidated from examining dynamics of the trade-offs. The revealed urban relational processes of city systems are at the core of the emerging science of cities and crucial for spatial planning and regional policymaking.

Most theorization of the spatial organization of city systems involves two key dimensions: hierarchy and proximity.[1–4] The former represents the hierarchical or 'vertical' relations of urban places and indicates the importance of cities, traditionally measured as their absolute size (e.g., population and GDP),[5,6] but increasingly measured through their centrality in terms of attractiveness in intercity interactions.[7,8] The latter represents the non-hierarchical or 'horizontal' relations that define city networks and indicates the interdependency among similar cities,[9,10] commonly gauged through a variety of functional and physical distance measures.[11–14] The hierarchy-dominated urban process, prominently identified in Central Place Theory (CPT), postulates that large central places have agglomeration economies in serving high-order goods and services over a wide hinterland, featuring asymmetric city-hinterland relations and competitions between central cities.[2,15,16] The proximity-dominated urban process, emphasized by city network theory, holds that city order is no longer the only determinant of urban functions and geographical proximity does not necessarily lead to interactions between cities, characterizing more symmetric and cooperation-based relations among similar cities. Practically, city systems are conceived as hierarchical networks that embed both hierarchy and proximity dimensions of external urban relations (Fig. 1a).

City size has been used as a common proxy to measure the hierarchy of city systems, which is well recognized as the rank-size law of city-size distribution.[5,6,17,18] Among these, the popular Gibrat's law argues that growth rates of individual cities are randomly distributed, leading to a power-law distribution for city size over time. These studies show that the scaling of city size as a hierarchy is consistent with various economic models that explain how systems grow through agglomeration.[17,19,20] However, this body of work generally suggests that the spatial organization (in essence the proximity between cities) among cities does not influence the distribution of city sizes. On the other hand, the celebrated CPT first showed that hinterlands around cities scaled across a geometric hierarchy in terms of their population size.[15] Along this city morphology perspective, enormous studies leveraging fractal geometry as a means have shown that urban patterns (both within and across cities) manifesting their self-similarity across different scales can be represented as spatial fractal structures with distinct hierarchical ordering that accords to the rank-size scaling.[21,22] They mostly build on the notion of development density decay with respect to distance in cities from their established center and postulate that similar underlying processes based on this notion operate across scales. Taken together, both bodies of existing work focus on the size of cities or density at locations, which essentially views city systems as 'space of places'.[23]

On the contrary, interactions between cities play a crucial role in the evolution of city systems,[24,25] featuring an interaction perspective that complies with the definition of 'space of flows' for contemporary cities.[7,8,23,26,27] Latest studies found that intercity population flows give rise to the power-law based city-size hierarchies that are not stationary.[28] These findings are theoretically grounded and related to the complexity of city systems, which postulates that the interplay between hierarchy and proximity gives rise to a latent structure of city systems.[2,24,29] In this sense, hierarchy and proximity are embedded in the underlying structures of city systems manifested by the complex networks of intercity flows and could be inferred by discovering the latent structures of city systems from intercity interactions. In sum, without an interaction perspective the space-of-places related approaches are subject to two limitations for a deep understanding of the underlying structures of city systems. First, the density-oriented city size and spatial distance (or travel time) are exogenous proxies and cannot faithfully represent the hierarchy and proximity dimensions, which are endogenous in the underlying structures of city systems and whose interplay must be accounted for in the measurement. Second, the latent space of the complex networks of intercity flows cannot be discovered, which deters the inference of hierarchy and proximity measures for cities.

Recent advances in network science show that complex networks with heterogeneous degree distributions and strong clustering have been conjectured to be embedded in hidden hyperbolic spaces, in which distances among nodes encode a balance between their similarity and popularity and, thus, determine their likelihood of being connected.[14,30–32] Hyperbolic space, a non-Euclidean metric space with constant negative curvature, is particularly well-suited for representing hierarchical relationships, such as those found in taxonomies, knowledge graphs, or tree-structured data.[33–35] It has been shown that embedding models able to find suitable underlying spaces offer geometric interpretations of the complex topologies observed in real networks, including scale-free degree distributions, the small-world effect, strong clustering, community structure and self-similarity, which are commonly found in human mobility networks.[30,33] Leveraging these advanced methods, we propose here a novel and theoretically grounded geometric model to unravel the interplay between hierarchy and proximity and to further examine the stationarity of the revealed structures of city systems with respect to such interplay. We hypothesize that hierarchy and proximity of cities and their interplay can be explained by assuming a latent hyperbolic geometry that reveals the underlying structures of city systems. This geometric approach undoubtably provides a coherent and comprehensive view to account for both hierarchy and proximity, hence leads to a superior modeling approach to those focusing on either one alone (e.g., various centrality measures and community detection methods). Our model uncovers, across countries, over time, and despite different mobility types, the bottom-up emergence of city hierarchies, along which the variations of city–hinterland relations are non-stationary in terms of their recursive manner and hinterland extent. These findings underscore a potential methodological paradigm shift for the science of cities, and carry practical implications for understanding the dynamics of city systems and informing regional and urban policy.

## A hyperbolic geometric model of city systems

We propose a geometric model of city systems by embedding intercity networks into the Poincaré ball space (Fig. 1a right). In Poincaré ball space,[35] a typical hyperbolic space with a constant curvature equal to -1, distances increase exponentially as points in the space approach the boundary (see 'Poincaré ball' in SI Methods). Notably, the root city of a tree structure, an analogy for hierarchical city systems, has the highest order and naturally resides near the origin while maintaining relatively short distances to all other cities on the tree. Cities at low orders are progressively pushed toward the boundary, and cities at different orders remain well-separated. In such a space, cities can be embedded in a polar coordinate system where the radial and angular dimensions encode hierarchy and proximity, respectively. The Poincaré distance between two cities $u, v$ with polar coordinates $(r_u, \theta_u)$ and $(r_v, \theta_v)$, can be approximated by:[33,35]

$$d(u,v) \approx r_u + r_v + 2\ln\left(sin\frac{\Delta\theta}{2}\right) \tag{1}$$

where $\Delta\theta = |\theta_u - \theta_v|$ represents the angular separation between $u$ and $v$. This equation reveals that the hyperbolic distance depends on the radii $r_u$ and $r_v$, which are latent measures of hierarchy for $u$ and $v$, and the angular separation $\Delta\theta$, which is the latent measure of proximity between $u$ and $v$. Reformulating (1) in exponential terms yields:

$$e^{-d(u,v)} \approx \frac{e^{-r_u}e^{-r_v}}{\sin^2\left(\frac{\Delta\theta}{2}\right)} \tag{2}$$

where $e^{-r_u}$ and $e^{-r_v}$ can be associated with the size of cities $u$ and $v$ and $sin^2(\frac{\Delta\theta}{2})$ is related to the distance between them, which together determines the mobility interaction strength between $u$ and $v$, and hence demonstrates how hierarchy and proximity measures in the latent space encode both the size and distance components in gravity models. Theoretically, hyperbolic geometry and gravity models are isomorphic.[32,36]

We first designed a sampling method based on weighted and directed random walks to sample sequences of nodes from intercity networks, which capture the topological, directional, and intensity features of interactions (Fig. 1b). Each intercity network is defined as a directed weighted graph $G = (V, E, W)$, where $V$ is a set of nodes (i.e., cities), $E$ is a set of directed edges, and $w_{u,v} \in W$ represents the weight (i.e., human movements) of an edge from node $u$ to $v$. Each random walk is defined as a sequence of nodes $s_1, s_2, \ldots, s_i$, where each transition between two consecutive nodes encodes the topological connections of networks and the direction of interactions. Collectively, high frequent transitions in the set of sampled walks reflect the intensity of interactions between nodes, as random walks are sampled according to predefined probability distributions based on the weights of edges. Given that a walk has just transited from node $u$ to $v$ at step $i$, the probability of transition from node $v$ to $t$ at step $i+1$ is defined as:

$$P(s_{i+1} = t | s_i = v, s_{i-1} = u) = \frac{w_{v,t}^{\alpha}[\beta \cdot \mathrm{I}(t = u) + (1-\beta) \cdot \mathrm{I}(t \neq u)]}{\sum_{p \in N(v)} w_{v,p}^{\alpha}[\beta \cdot \mathrm{I}(p = u) + (1-\beta) \cdot \mathrm{I}(p \neq u)]} \tag{3}$$

where $\mathrm{I}(\cdot)$ is a switch function that outputs binary values, $N(v)$ denotes the set of nodes directly reachable from $v$, the parameter $\alpha$ controls the preference for higher-weighted edges (i.e., edges with stronger connectivity), and the parameter $\beta$ adjusts the tendency to revisit the previously visited node to avoid walks getting trapped into local neighborhoods.

We then fitted our geometric model with the sampled sequences of transitions to infer latent measures of hierarchy and proximity for cities while imposing meaningful constraints to the hierarchy measure (Fig. 1b). To achieve this, we adopt a contrastive learning approach with regularization.[35,37] Formally, we minimize the loss function:

$$L = -\sum_{(u,v)\in D^+} log(\frac{e^{-d(u,v)}}{e^{-d(u,v)} + \sum_{v'\in D_u^-} e^{-d(u,v')}}) + \lambda \cdot \max_{v\in V}(r_v) \quad (4)$$

where $\lambda$ is a hyperparameter that controls the strength of the regularization. The first term in (4) can be interpreted as a soft ranking loss where the hyperbolic distance $d(u,v)$ of observed transition pair $(u,v)$ in the set $D^+$ of sampled walks should be shorter than that of pair $(u,v')$ from the randomly generated set $D_u^-$ of negative samples. It also ensures that $v'$ is orderly rather than arbitrarily far apart according to their distance $d(u,v')$. The second term implicitly assumes that the hierarchy measure of cities has a lower bound, meaning that the largest radial coordinate of cities is subject to an upper bound. It creates a necessary condition for the distribution of the hierarchy measure to converge to a power-law distribution even though it is not a sufficient one,[38] which facilitates the subsequential examination of city hierarchies, given that the commonly reported rank-size law of city systems.[8]

**Emergent hierarchies of city systems**

We fit the hyperbolic geometric model to 12 different datasets including two types of intercity human mobility data, i.e., daily-travel (short-term) and annual-migration (long-term), from China and United States for years of 2019~2021: D1 and D3 are the daily-travel datasets for China and the US, respectively; D2 and D4 are the annual-migration datasets for China (years of 2019~2023) and the US, respectively (see 'Data and pre-processing' in SI Methods). Notably, all datasets cover the years before and after the outbreak of Covid-19. The embedding results for different datasets all show an emergent order of cities. For example, the results for China 2019 data in D1 (Fig. 1c) depict that national centers (e.g., Beijing and Shanghai) are located close to the origin, followed by provincial centers (e.g., Nanjing and Xi'an), then the other local centers and smaller cities along the radial axis. Along the angular axis, cities in the same province are clustered in closer proximity. In addition, the inferred hierarchy measure is highly correlated with the natural logarithm of city's population and GDP (Fig. 2). It suggests that the hierarchy measure is the latent variable representing the overall attractiveness of a city owing to its level of urban functions, which is closely associated with city size. Along with this reasoning, the rank-size law is further tested given this latent relationship between the hierarchy measure and the city size. Fig. 2 shows that distributions of hierarchy measures for all datasets exhibit power law behavior. Specifically, they follow the double pareto distributions, which have essentially a lognormal body but a power law distribution in the tail.[38]

Given the emergent hierarchy of cities characterized by the hierarchy measure, we postulate that the interplay between hierarchy and proximity can be understood by examining the city-hinterland relation that a larger (high-order) city serves as the center providing high-order goods and services to a set of smaller (low-order) cities at different orders making up the hinterland or market area. Similar as a branching tree structure, this relation is recursive in that smaller cities may also be centers for nested hinterlands comprising even smaller cities. By evaluating the variation of city-hinterland relations in the latent structure of city systems, the two urban relational processes, i.e., hierarchy- and proximity-dominated, can be elucidated. Specifically, we define that a city-hinterland structure is order dependent such that a central city links to low-order cities at only one certain order in the hinterland (see 'City-hinterland structures' in SI Methods). Therefore, a central city may have city-hinterland structures at multiple orders, collectively making up a nested hierarchy of cities. Analytical derivations (see 'Hierarchy measure and city order' in SI Methods) show that the hierarchy measure is in essence a continuous representation of city order in CPT. Given these definitions, we extracted the tree structures of city hierarchies based on two fundamental rules (see 'Tree construction algorithm' in SI Methods): 1) cities at the same order do not become one another's hinterland, while a low-order city (child node) chooses only one high-order city (parent node) to become its hinterland; 2) the city-hinterland relation minimizes the Poincaré distance of links (equivalently maximizing the interaction strength) between low-order cities and a high-order city. The resulting tree structures (Fig. S1a) contain levels of cities classified into discrete city orders according to their hierarchy measure, which are consistent with the commonly recognized city ranking in practice (Fig. S2).

This nesting property of the city-hinterland relation reflects the rank-size law in the city-hinterland structure, which can be measured by the nesting factor (NF) defined as one plus the number of low-order cities served by a central city in a city-hinterland structure (see 'Nesting factor' in SI Methods). Another important property of a city-hinterland structure is its range, which delineates the boundary of the hinterland in terms of proximity between the central city and the low-order city at the boundary. Specifically, it refers to the most 'distant' proximity that a low-order city in a city-hinterland structure is willing to bear while reaching a threshold level of interaction strength (the interaction threshold) with the central city required to become its hinterland (see 'Hinterland range' in SI Methods). According to Christaller's CPT,[2] the NF is fixed for a particular urban system, e.g., market principle with NF=3, and the range has a fixed functional relationship with the NF.

**Non-stationary city-hinterland structures along city hierarchies**

We further examine whether the extracted city-hinterland relations in the latent structure of city systems are stationary with respect to the NF and the range properties. We find that the highly skewed distributions of NF can be characterized by relative stable medians of NF, i.e., 3 and 2 for short-term data of China and the US, respectively (Fig. 3a,e). For the long-term data, in the year of Covid-19 outbreak the medians of NF evidently decrease (from 3 to 2 for both China and the US), while in the normal years they are stable at 3 (Fig. 3a,e). That said, we further find that NF's means and variations by order increase as the city order gets lower (larger number) for both the short-term and long-term data of China (Fig. 3b). This implies that city-hinterland structures from the high to low orders of the city hierarchies tend to serve more than the median number (>3) of low-order cities. Additionally, results for the long-term datasets show more complex patterns, where the mean and variation of NF over time are dramatically larger at orders 4-5 and 8-9. These non-stationary variations of NF for China are further confirmed by results in Fig. S3, which find that, compared to those at the same order, city-hinterland structures with NF larger than the median value tend to have their central cities at higher orders. Moreover, these city-hinterland structures are mostly found at lower orders for all datasets (Fig. S4), which indicates that small cities at the bottom of city hierarchies are more likely to be attracted by large cities with high-order functions even though they tend to have relatively low interaction thresholds (large Poincaré distances) with these large cities. In contrast, while NF's means by order are varying without a clear trend for both the short-term and long-term data of the US, they show significant differences at high orders and converge at lower orders over time (Fig. 3f). This indicates that the nesting property of the city-hinterland relation is relatively stable over time at low orders of the city hierarchies, while changing dramatically at high orders. We argue that in the latent structure of city systems city-hinterland relations in terms of the nesting property is evidently non-stationary conditioned on the city order regardless of country, mobility type, or time, even though their NFs have characteristic values (medians of NF in the range of [2, 3]) that are relatively stable over the three dimensions.

We find that hinterland ranges have a general descending trend as moving down the city order for short-term datasets of both China and the US (Fig. 3c,g and Fig. S5). The slop of the fitted line (orange) is theoretically linked to a uniform nesting factor if the stationary city hierarchy is assumed (see 'Inferring a global $K$ from hinterland ranges' in SI Methods). The estimated uniform NFs are in the interval [1.85, 3.07] for China and [2.18, 3.62] for the US, respectively (Fig. S5). While slightly deviating the corresponding median value of NF for each country and year, they are roughly consistent with the value interval [2, 3] of these characteristic NFs, which implies certain intrinsic coherence of these results. Notably, low R-squares of model fitting in Fig. S5 signal large proportions of range variation unexplained by the linear relationships with order, and model fittings on the China datasets have generally lower R-squares than those on the US datasets. In Fig. S6, the former is further confirmed by the non-stationary variations of residuals by city order from the fitted models, and the latter can be attributed to 1) the larger variation of residuals especially at the high orders and 2) the larger deviation of concentration of residuals from zero, which implies that City-hinterland structures in China exhibit more non-stationarity in terms of hinterland range than those in the US. Despite that, we did not find a consistent relationship between the variation of range and that of NF by city order across all datasets (Fig. S6), as the range is not solely determined by the NF (also see 'Inferring a global $K$ from hinterland ranges' in SI Methods).

Additionally, we find that non-stationary concentrations of ranges distributed across city orders indicate characteristic sizes, such as $3° \pm 1°$, $15° \pm 5°$, and $60° \pm 5°$ (Fig. 3c), which correspond to geographic scales of metropolitan areas, urban-regional agglomerations, and regional divisions, respectively, from the short-term datasets of China (Fig. 3d excluding the lower-right panel). Equivalent geographic scales, i.e., MSA, state, and Census Division (Fig. 3h) can be identified as well by the characteristic range sizes: $5° \pm 5°$, $20° \pm 5°$, and $55° \pm 5°$, respectively (Fig. 3g,h), from the US short-term datasets. Systematic validations of such correspondence are detailed in 'Characteristic ranges and geographic scales' section in SI Methods and the results shown in Fig. S7d. We argue that for daily-travel city systems city-hinterland relations in terms of hinterland range indicate spatially non-stationary extents of 'market areas' for central cities along the city order, which are confined by geographic or administrative boundaries at different scales. In contrast, the correspondence between ranges of city-hinterland structures and geographic scales does not exist for long-term datasets of both China and the US (Fig. 3d lower-right panel). These results imply that long-term migrations are not constrained by geographic or administrative boundaries, but they can be greatly affected by the disruption of Covid-19.

**Trade-offs between city hierarchy and hinterland range**

We argue that the non-stationary structures of city hierarchies originate from optimizing trade-offs between central city's hierarchy and the range of hinterlands in determining the interaction thresholds of mobility for city-hinterland structures (Equation 2). To demonstrate this, we derive isolines of interaction thresholds for all city-hinterland structures on the city hierarchies for each dataset (see 'Deriving isolines of interaction thresholds' in SI Methods). Fig. 4 and Fig. S8 show that such trade-offs are represented by the general negative relationships between city hierarchy and hinterland range across all datasets. If stationary NFs and ranges of city-hinterland structures are assumed, the isolines of interaction thresholds would be perfectly parallel straight lines with negative slops (see Supplementary Note 2). Conversely, empirical isolines in Fig. 4 that are far apart from parallel straight lines and vary with city orders reflect dynamics of trade-offs between city hierarchy and hinterland range, which indeed give rise to the non-stationary city-hinterland structures. For instance, on one of three trees for 2019 short-term data of China whose root node is Shanghai, central cities at order 2 show very small ranges (e.g., Suzhou-Wuxi and Hangzhou-Jinhua), which features very strong interactions with proximate low-order cities (Fig. 4a). The same holds for the Cincinnati-Columbus at order 2 of the Indianapolis tree for 2021 short-term data of the US (Fig. 4c). As these city-hinterland structures have higher interaction thresholds (smaller Poincaré distance), a shorter range is required to sustain their hinterlands given their hierarchy fixed. In other words, the hierarchy dominates the trade-off with a smaller range. Conversely, order-3 cities on the Shanghai tree have larger ranges, which involves the cross-region interactions with remote hinterlands, (e.g., Wuhan-Ji'an and Wuxi-Xuzhou in Fig. 4a). The same holds true for Minneapolis-Carson City at order 3 of the Indianapolis tree for 2021 short-term data of the US (Fig. 4c). As they have lower interaction thresholds (larger Poincaré distance), a larger range is necessary to sustain their hinterlands given their hierarchy fixed, which means that the range (proximity) relatively dominates the trade-off.

For 2019 daily-travel datasets (Fig. 4e and Fig. S9), moving up the city order cities in China have their hinterland ranges increased more than proportionately, which can be indicated by a negative Change Rate of Interaction Thresholds (CRIT, see 'Change Rate of Interaction Thresholds' in SI Methods), while cities in the US have their ranges increased less than proportionately, indicated by a positive CRIT. In other words, the proximity is more likely to dominate the trade-off for higher-order cities in China, hence implying a relatively stronger proximity-dominated urban process, while the hierarchy is more likely to dominate the trade-offs for higher-order cities in the US involving a relatively stronger hierarchy-dominated urban process. For 2019 long-term data of China, the interaction thresholds for cities to sustain their hinterlands first increase at orders 2 to 4 and then decrease as moving down the order (Fig. S9). This indicates that hinterland ranges enlarge less than proportionately when moving up the city order (positive CRIT values in Fig. 4e) and implies that higher-order cities have a relatively stronger hierarchy-dominated urban process in contrast to what shown for the short-term datasets. It is noteworthy that cities at orders 2 to 4, exhibiting strong interactions with their proximate areas, are mostly provincial capitals or regional centers, which usually attract most migrants within provinces (e.g., Chengdu-Yibin and Kunming-Shaotong in Fig. 4b). The result of 2019 long-term data for the US shows a similar pattern (Fig. 4d) that moving up the city order hierarchy tends to dominate the trade-offs (positive CRIT measures in Fig. 4e), hence implying a relatively stronger hierarchy-dominated urban process for higher-order cities.

Over time, the proximity-dominated urban process generally becomes weaker for short-term datasets of China due to the Covid-19 related travel restrictions. Specifically, strong proximity-dominated urban process for Beijing tree becomes weak hierarchy-dominated urban process after 2019, while weak proximity-dominated urban process for Chongqing tree becomes stronger since 2020. For long-term datasets of China, the hierarchy-dominated urban process was weakened by the impact of Covid-19 related non-pharmaceutical interventions not only on travel but also on economic activities.[39,40] Consistently, Chongqing tree has a stronger hierarchy-dominated urban process than Hangzhou tree does even though both trees have structural changes (Fig. S10). Due to Covid-19 impacts since 2020, the hierarchy-dominated urban process with respect to short-term datasets of the US becomes stronger and then weaker, while the hierarchy-dominated urban process with respect to long-term datasets of the US exhibits a stronger trend. They both show complex variations across trees of city-hinterland structures due to significant structural changes (Fig. S11).

## Discussion

In this Article, we build a hyperbolic geometric model for city networks of human mobility that unravel the latent hierarchical structures of a system of spatially interactive cities from a view of human mobility. This is done in the model by explicitly inferring the measures of hierarchy and proximity, two fundamental dimensions with micro-foundations rooted in urban system theories, and accounting for the trade-off between them. This model provides a bottom-up approach that shows the emergence of global structures of city hierarchies purely based on the spatial interactions between paired cities, which features a self-organizing criticality of complex systems. This contrasts to a top-down approach that presumes information about the hierarchy is known, such as the PSO model,[31] and infers pairwise proximity. Our model has been found robust across all datasets and shows superior performance on most datasets (see Supplementary Note 1, Tables S1, S2 and S3). The hierarchy measure as the latent variable for city size provides a strong counterevidence against what Gibrat's law or random growth models have suggested, i.e., spatial relations among cities do not influence the distribution of city size.[7,8,26,41] Instead, it confirms the findings of recent studies on spatial-grouping relation (larger cities tend to serve as centers around which smaller cities are grouped) and migration flows between cities of different sizes do indeed play a crucial role in shaping the landscape of city distribution.[17]

More importantly, our finding shows that non-stationarity of city-hinterland relations holds for all city systems investigated. Firstly, there is not a universal scaling factor for the nesting property of city-hinterland relations, which is revealed by the varying nesting factors conditioned on the city order while they have characteristic values that are relatively stable and roughly consistent with the market principle of CPT. This finding extends the recent studies on spatial-grouping property and central place property that either assumes a fixed global nesting factor or a homogeneous distribution of demands.[17,42] These studies in essence restrict the variability of central place property at the micro-level for city-hinterland relations and, thus, leads to a biased identification of spatial fractal structures in the recursively nested hinterlands. Secondly, we find that proximity-based hinterland range decreases when moving down the city order and these relationships show certain intrinsic coherence with the characteristic values of NF found previously. However, they show significant non-stationarity for daily-travel urban systems. This finding extends recent studies on testing the spacing-out property of cities at the same order, which entirely rely on road distance-based Voronoi partition and homogeneous population distribution for hinterland delineation and range determination.[27,43] Consequently, they are unable to capture the heterogeneity of range distribution and the delineated hinterland geographies. More importantly, we find that hinterland ranges have characteristic sizes corresponding to the geographic boundaries of urbanization at multiple spatial scales for both China and the US. It confirms the findings of recent study on the characteristic spatial scales of human mobility,[44] and unveils the human behavioral dimension of hinterland ranges that are essentially a socioeconomic construction beyond the constraints imposed by physical barriers, administrative boundaries, and even road infrastructures.

Our results reveal that the non-stationary city-hinterland structures of city systems originate from optimizing trade-offs between city hierarchy and hinterland range. Hierarchy- and proximity-dominated urban processes have been elucidated from the dynamics of the trade-offs in different city systems. Except the daily-travel city systems of China exhibiting proximity-dominated urban process, all others show hierarchy-dominated urban processes. Except the strengthened urban process in annual-migration city systems of the US, urban processes in other city systems are all weakened since the Covid-19 outbreak in 2020. This approach echoes the urban scaling law wherein the exponent (greater or smaller than 1) in a power function of city size indicates the level of agglomeration effects,[26,45–48] while complementing it by also accounting for the network effect (proximity-dominated urban process). Remarkably, isolines of interaction thresholds depicting trade-offs in this approach are in line with the behavioral isolines revealed by space-preference functions to address the trade-off between travel distance and town population for retail opportunities faced by shipping customers.[29,49,50] The two approaches are equivalent in both deriving a global structure (hierarchical city systems vs. space preference) from local-level interactions (intercity mobility vs. individual preference), while our approach derives the latent measures of hierarchy and proximity purely from interaction data without relying on additional correlates such as population and distance.

The evaluation of the latent measures for hierarchy and proximity centers on the extracted tree structures of city-hinterland relations in city systems and neglects the interactions between cities at the same order and the possibility of two central cities sharing their market areas (Fig. S1b). Besides, we focus on the spatial organization of city systems without considering the temporal dimension of city growth. These aspects could be incorporated in future studies. This geometric approach can be recognized as a unifying framework that has been found to connect with most mainstream urban and regional theories. It can be applied to study human settlement systems with spatial interactions at a range of scales, from internal urban structures to city systems within a country, and to the world trade networks between countries,[12,41] as long as they exhibit heterogeneity and clustering of interactions. From evaluating CPT's hypothetic hexagon geometry to examining the fractal geometry of urban density patterns and to unveiling the hidden hyperbolic geometry of urban interactions, it may suggest a methodological paradigm shift for the science of cities, calling for a transition from Euclidean space to non-Euclidean space, from static and absolute space to relational and relative space, and from space of places to space of flows.

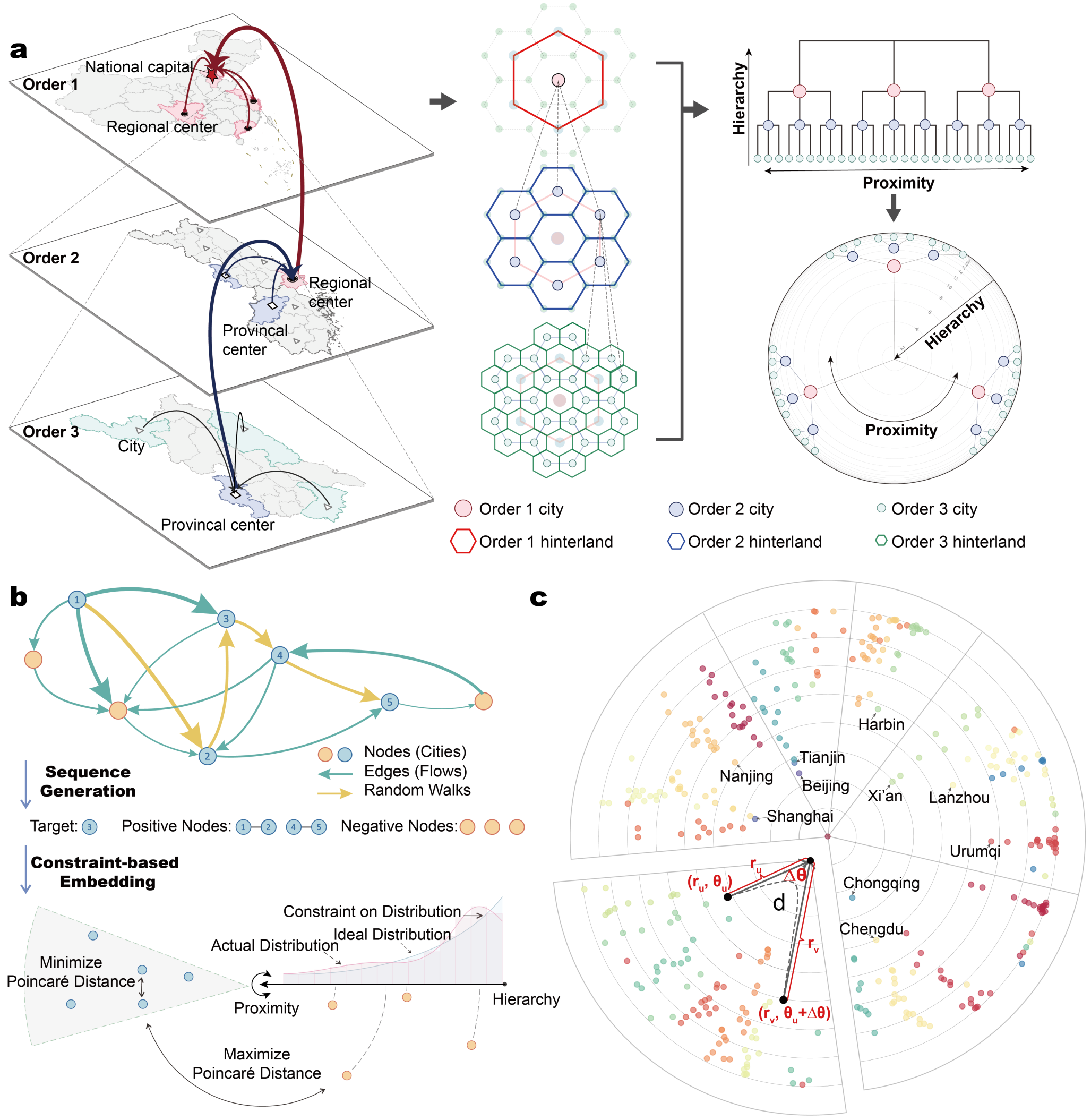

**Fig. 1. Hierarchical networks of city systems and hyperbolic embeddings. a,** Hierarchical networks of city systems are illustrated (left), where high-order cities (e.g., the national capital and regional centers) and low-order cities (e.g., provincial centers and normal cities) forming nested city-hinterland organizations (middle) through linkages of human movements. Such hierarchical networks of city systems can be represented as tree structures, whose 'vertical' and 'horizontal' relations are represented by hierarchy and proximity dimensions, respectively (right). The two dimensions can be embedded in a Poincaré disk space, where the radial and angular dimensions encode hierarchy and proximity, respectively. **b,** Weighted and directed random walks (yellow arrows) are sampled to capture the topological, directional, and intensity features of intercity interactions and then geometric models are fitted to infer constraint-based embeddings of cities. **c,** Example embeddings of cities from China's short-term mobility dataset show that high-order cities are around the centroid and low-order cities are near the periphery.

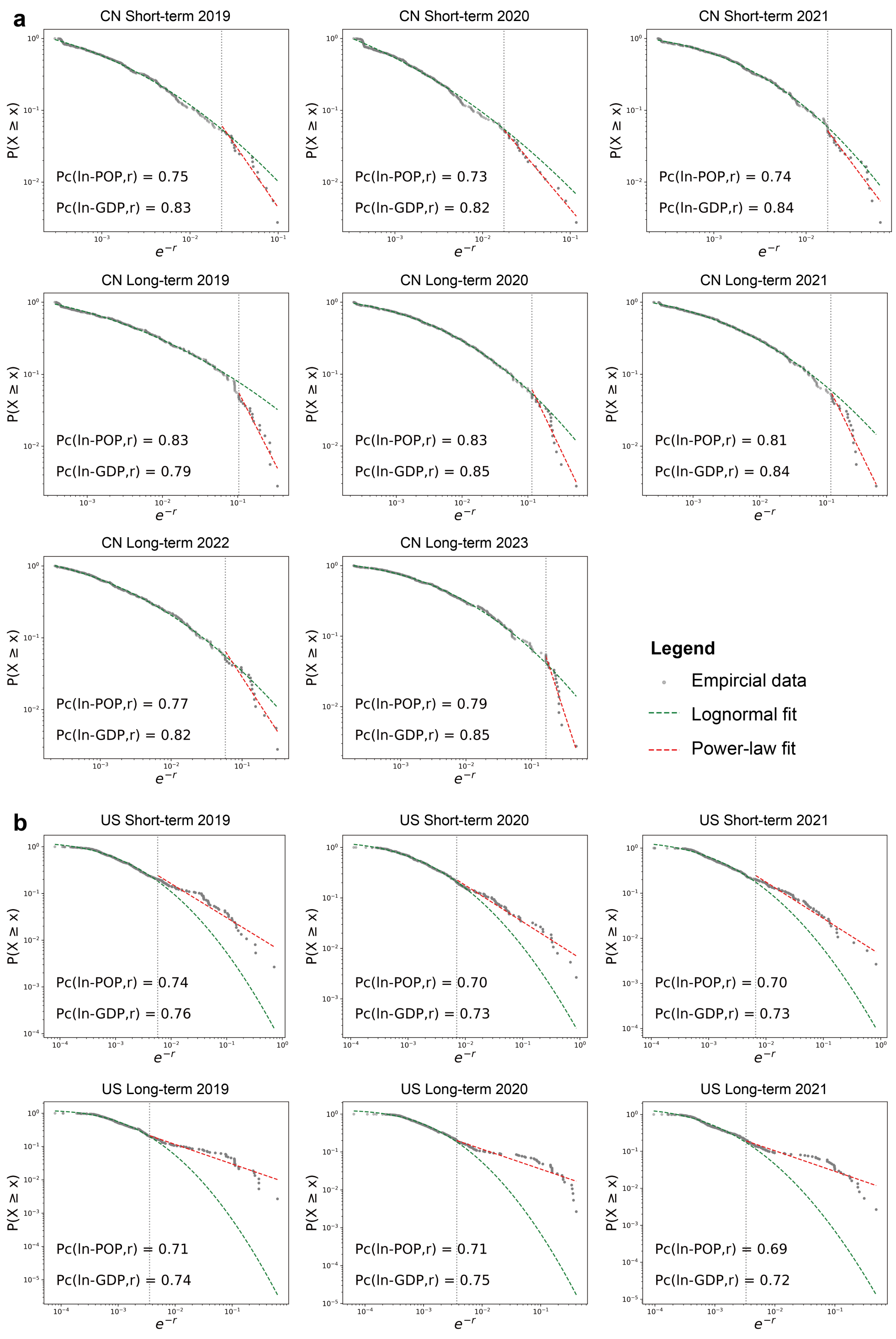

**Fig. 2. Complementary cumulative distributions (CCD) of the latent measure $e^{-r}$ underlying city size and its correlation with city population and GDP.** $r$ is the hierarchy measure and $e^{-r}$ is equivalent to city size according to Equation 2. Pc is the Pearson correlation coefficient. ln-POP and ln-GDP are the natural logarithm of the annual population and GDP of cities, respectively. The CCDs of $e^{-r}$ are to evaluate how well they follow the rank-size law. Empirical data (scatter points) represent observed distributions of $e^{-r}$, and the power-law (dashed red lines) and lognormal functions (dashed green lines) are also fitted with the datasets for China (**a**) and the US (**b**), respectively. Gray vertical dashed lines mark truncation starts for power-law fitting:

the 95th percentile of $e^{-r}$ for datasets from China, and the 80th percentile for datasets from the United States. Results show that the $e^{-r}$ follow the double pareto distributions, which have essentially a lognormal body but a power law distribution in the tail.

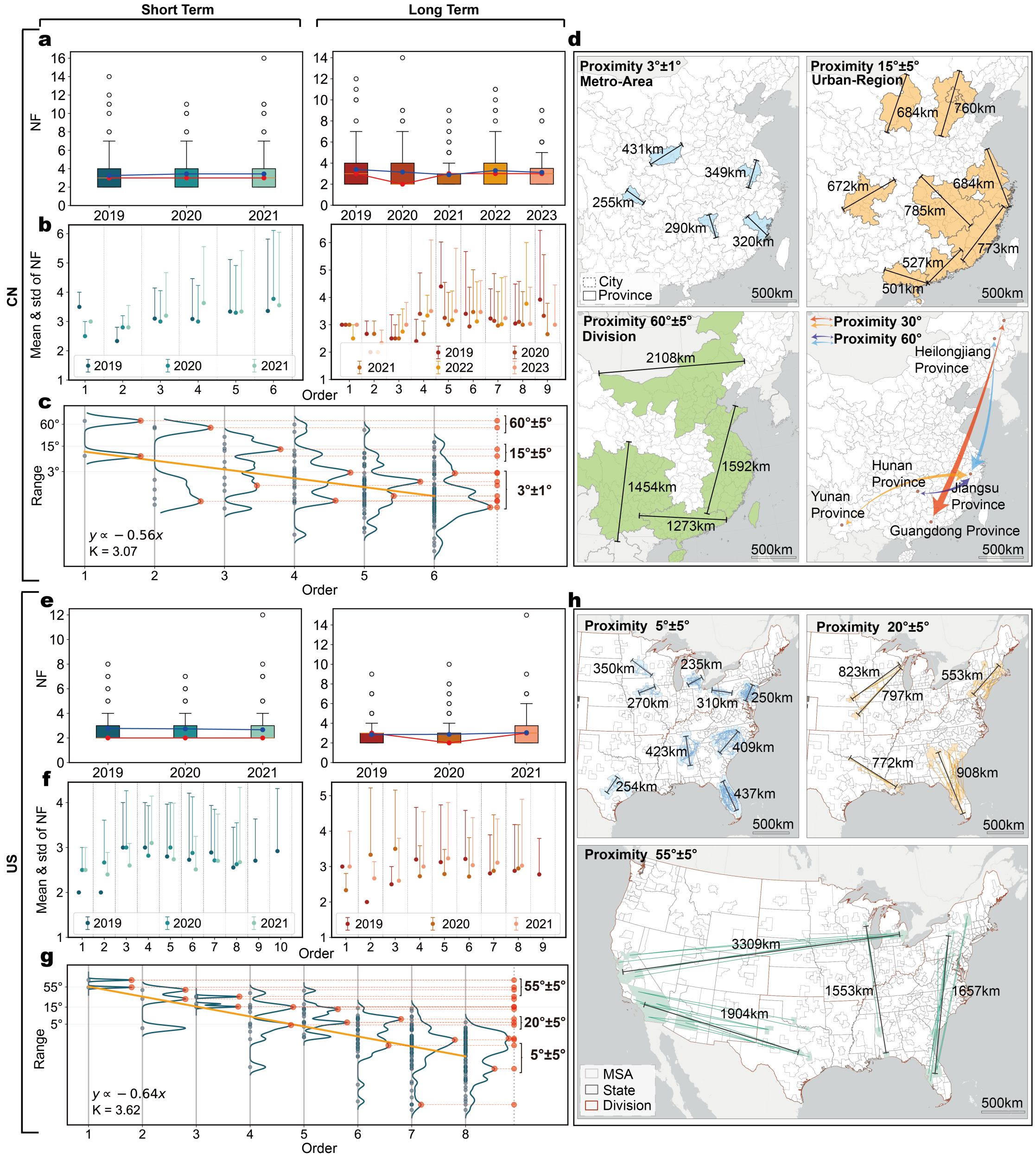

**Fig. 3. Nesting factor (NF) and range of city-hinterland structures in city systems.** Distribution of NF for short-term and long-term datasets of China (**a**) and the US (**e**). Red (blue) dot and line represent the median (mean) of NF to highlight its variation over time. Mean and standard deviation (std) of NF by city order for short-term and long-term datasets of China (**b**) and the US (**f**). Distribution of hinterland range measured by proximity by city order for short-term dataset of China 2019 (**c**) and the US 2020 (**g**). Orange lines indicate linear model fittings and global $K$ values are estimated based on the slope of fitted lines. Characteristic range sizes can be identified for short-term datasets of China (**d**) and the US (**h**) correspond to different geographic scales of urbanizations. Range sizes for long-term datasets do not correspond to geographic scales for both countries (**d** lower-right).

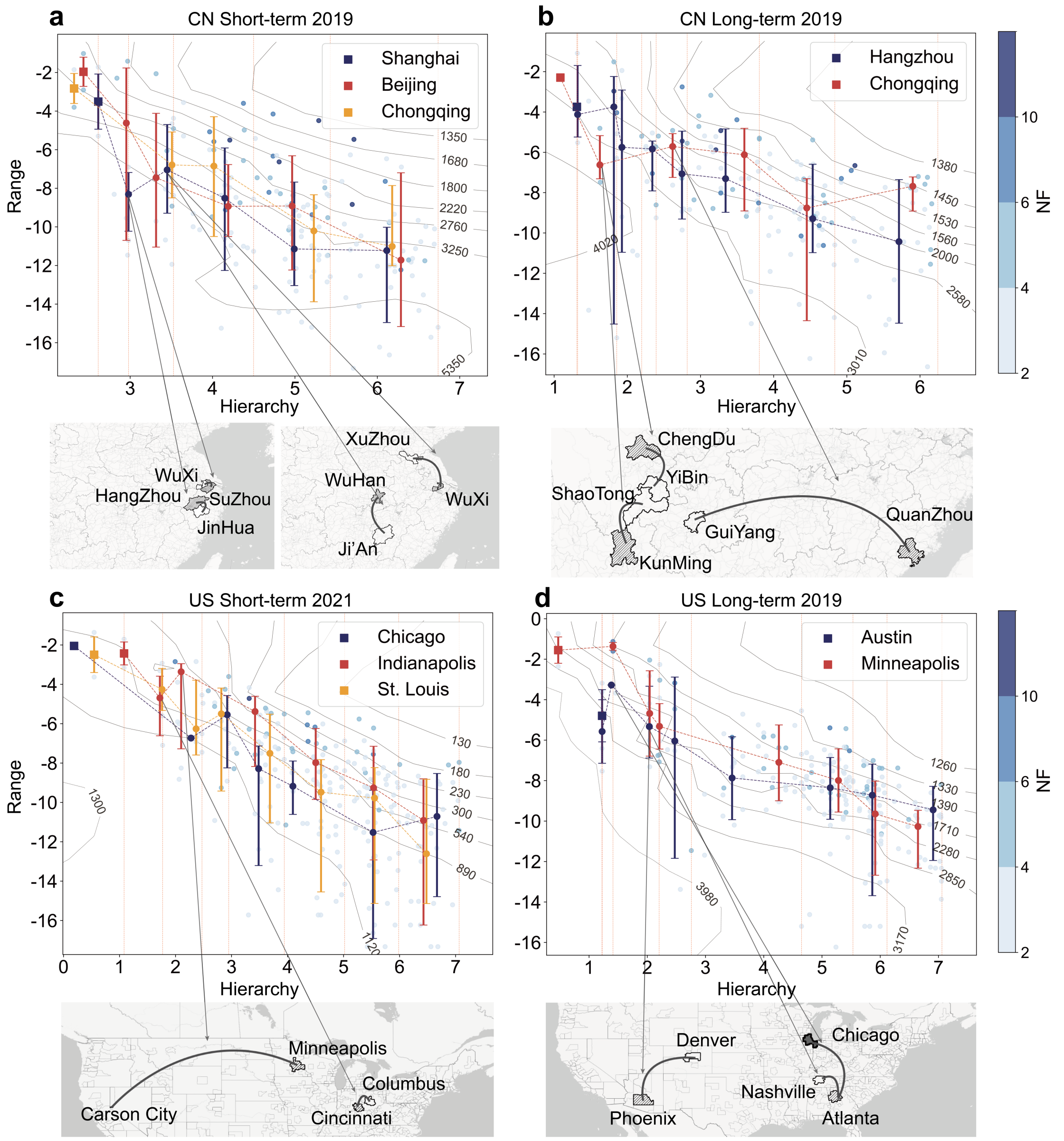

e

| CRIT | 2019 | | 2020 | | 2021 | |
|---|---|---|---|---|---|---|
| CN Short-term | *All* | *-0.305* | *All* | *-0.185* | *All* | *-0.232* |
| | Shanghai | -0.126 | Shanghai | -0.089 | Shanghai | 0.046 |
| | Beijing | -0.332 | Beijing | 0.007 | Beijing | 0.078 |
| | Chongqing | -0.025 | Chongqing | -0.499 | Chongqing | -0.299 |
| CN Long-term | *All* | *0.355* | *All* | *0.301* | *All* | *0.317* |
| | Hangzhou | 0.277 | Hangzhou | 0.204 | Shanghai | 0.282 |
| | Chongqing | 0.701 | Chongqing | 0.447 | Chongqing | 0.328 |
| US Short-term | *All* | *0.217* | *All* | *0.241* | *All* | *0.161* |
| | Chicago | 0.322 | Chicago | 0.098 | Chicago | 0.310 |
| | Las Vegas | 0.264 | Minneapolis | -0.018 | Indianapolis | 0.217 |
| | Washington | 0.125 | St. Louis | 0.372 | St. Louis | 0.082 |
| US Long-term | *All* | *0.410* | *All* | *0.500* | *All* | *0.485* |
| | Austin | 0.526 | Austin | 1.154 | Chicago | 0.563 |
| | Minneapolis | 0.043 | Chicago | 0.611 | Minneapolis | 0.243 |
| | | | Minneapolis | -0.108 | Nashville | 0.234 |
| | | | St. Louis | 0.402 | | |

**Fig. 4. Dynamics of trade-offs between city hierarchy and hinterland range. a~d,** Distributions of hinterland ranges for central cities by city order for different trees extracted for each dataset. Results of four datasets are showcased here and complete

results for all datasets can be found in SI Appendix, Fig. S8. Gray dash lines discretize the hierarchy measure into city orders. Trade-offs are indicated by large dots, whose coordinates are the median of hierarchy measures of central cities at each order versus the median of their hinterland ranges. Error bars represent the 10th–90th percentile of the distribution of ranges. Square dots are the root cities (city names labeled in legend) for extracted trees. Dots are linked by dashed lines for each tree. If there is only one central city at an order, its name is annotated. Gray lines represent derived isolines of interaction thresholds with annotations indicating the estimated mobility strength, which facilitate temporal comparisons. Isolines are derived based on the interaction thresholds of all city-hinterland structures represented by small dots in the background with color gradients indicating their nesting factor (NF) values. Dynamics of trade-offs are identified and showcased by pairs of central and hinterland cities illustrated as locations linked by curves on maps. The locations of central cities are shaded. **e,** The Change Rate of Interaction Thresholds (CRIT) measures the change of trade-offs by city order, which are calculated for the overall average (*All*) and different trees (root city name) for each dataset.

## References

1. H. M. Abdel-Rahman, A. Anas, "Theories of systems of cities" in Handbook of Regional and Urban Economics, (Elsevier, 2004), pp. 2293–2339.
2. G. F. Mulligan, M. D. Partridge, J. I. Carruthers, Central place theory and its reemergence in regional science. Ann Reg Sci 48, 405–431 (2012).
3. P. J. Taylor, M. Hoyler, R. Verbruggen, External urban relational process: Introducing central flow theory to complement central place theory. Urban Stud. 47, 2803–2818 (2010).
4. M. Batty, The new science of cities (MIT press, 2013).
5. M. Batty, Rank clocks. Nat 444, 592–596 (2006).
6. L. M. Bettencourt, The origins of scaling in cities. Sci 340, 1438–1441 (2013).
7. A. Bassolas, et al., Hierarchical organization of urban mobility and its connection with city livability. Nat. Commun. 10, 4817 (2019).
8. Y. Xu, et al., Urban dynamics through the lens of human mobility. Nat. Comput. Sci. 3, 611–620 (2023).
9. R. Capello, The city network paradigm: measuring urban network externalities. Urban Stud. 37, 1925–1945 (2000).
10. B. Derudder, P. J. Taylor, Central flow theory: Comparative connectivities in the world-city network. Reg. Stud. 52, 1029–1040 (2018).
11. P. J. Taylor, Specification of the world city network. Geogr. Anal. 33, 181–194 (2001).
12. G. García-Pérez, M. Boguñá, A. Allard, M. Á. Serrano, The hidden hyperbolic geometry of international trade: World Trade Atlas 1870–2013. Sci Rep 6, 33441 (2016).
13. E. Arcaute, et al., Cities and regions in Britain through hierarchical percolation. R. Soc. Open Sci. 3, 150691 (2016).
14. D. Brockmann, D. Helbing, The hidden geometry of complex, network-driven contagion phenomena. Sci 342, 1337–1342 (2013).

15. H. Beguin, Christaller's central place postulates: a commentary. Ann Reg Sci 26, 209–229 (1992).
16. T. Tabuchi, J.-F. Thisse, A new economic geography model of central places. J. Urban Econ. 69, 240–252 (2011).
17. X. Gabaix, Zipf's law for cities: an explanation. Q J Econ 114, 739–767 (1999).
18. X. Gabaix, Y. M. Ioannides, "The evolution of city size distributions" in Handbook of Regional and Urban Economics, (Elsevier, 2004), pp. 2341–2378.
19. G. Duranton, Some foundations for Zipf's law: Product proliferation and local spillovers. Reg. Sci. Urban Econ. 36, 542–563 (2006).
20. E. Rossi-Hansberg, M. L. Wright, Urban structure and growth. Rev Econ Stud 74, 597–624 (2007).
21. M. Batty, The size, scale, and shape of cities. Sci 319, 769–771 (2008).
22. H. A. Makse, S. Havlin, H. E. Stanley, Modelling urban growth patterns. Nat 377, 608–612 (1995).
23. M. Castells, The rise of the network society (John wiley & sons, 2011).
24. D. Pumain, "Alternative explanations of hierarchical differentiation in urban systems" in Hierarchy in Natural and Social Sciences, (Springer, 2006), pp. 169–222.
25. E. J. Meijers, M. J. Burger, M. M. Hoogerbrugge, Borrowing size in networks of cities: City size, network connectivity and metropolitan functions in Europe. Papers in regional science 95, 181–199 (2016).
26. T. Louail, et al., Uncovering the spatial structure of mobility networks. Nat. Commun. 6, 6007 (2015).
27. L. Pappalardo, E. Manley, V. Sekara, L. Alessandretti, Future directions in human mobility science. Nat. Comput. Sci. 3, 588–600 (2023).
28. S. M. Reia, P. S. C. Rao, M. Barthelemy, S. V. Ukkusuri, Domestic migration and city rank dynamics. Nat. Cities 2, 38–46 (2025).
29. G. Rushton, Analysis of Spatial Behavior by Revealed Space Preference. Ann. Assoc. Am. Geogr. 59, 391–400 (1969).
30. M. Boguna, et al., Network geometry. Nat. Rev. Phys. 3, 114–135 (2021).
31. F. Papadopoulos, M. Kitsak, M. Á. Serrano, M. Boguná, D. Krioukov, Popularity versus similarity in growing networks. Nat 489, 537–540 (2012).
32. M. Boguná, F. Papadopoulos, D. Krioukov, Sustaining the internet with hyperbolic mapping. Nat. Commun. 1, 62 (2010).
33. D. Krioukov, F. Papadopoulos, M. Kitsak, A. Vahdat, M. Boguná, Hyperbolic geometry of complex networks. Phys. Rev. E 82, 036106 (2010).
34. M. Gromov, "Hyperbolic groups" in Essays in Group Theory, (Springer, 1987), pp. 75–263.
35. M. Nickel, D. Kiela, Poincaré embeddings for learning hierarchical representations. Adv Neural Inf Process Syst 30 (2017).
36. D. Krioukov, F. Papadopoulos, A. Vahdat, M. Boguná, Curvature and temperature of complex networks. Phys. Rev. E 80, 035101 (2009).
37. A. Radford, et al., Learning transferable visual models from natural language supervision in International Conference on Machine Learning, (PmLR, 2021), pp.

8748–8763.
38. M. Mitzenmacher, A brief history of generative models for power law and lognormal distributions. Internet Math. 1, 226–251 (2004).
39. M. Pangallo, et al., The unequal effects of the health–economy trade-off during the COVID-19 pandemic. Nat. Hum. Behav. 8, 264–275 (2024).
40. D. J. Haw, et al., Optimizing social and economic activity while containing SARS-CoV-2 transmission using DAEDALUS. Nat. Comput. Sci. 2, 223–233 (2022).
41. M. Batty, P. A. Longley, Fractal cities: a geometry of form and function (Academic press, 1994).
42. W.-T. Hsu, Central place theory and city size distribution. Econ. J. 122, 903–932 (2012).
43. T. Louail, M. Barthelemy, A dominance tree approach to systems of cities. Comput. Environ. Urban Syst. 97, 101856 (2022).
44. L. Alessandretti, U. Aslak, S. Lehmann, The scales of human mobility. Nat 587, 402–407 (2020).
45. L. M. Bettencourt, J. Lobo, D. Helbing, C. Kühnert, G. B. West, Growth, innovation, scaling, and the pace of life in cities. Proc. Natl. Acad. Sci. 104, 7301–7306 (2007).
46. R. Li, et al., Simple spatial scaling rules behind complex cities. Nat. Commun. 8, 1841 (2017).
47. S. Sarkar, et al., Evidence for localization and urbanization economies in urban scaling. R. Soc. Open Sci. 7, 191638 (2020).
48. D. Pumain, F. Paulus, C. Vacchiani-Marcuzzo, J. Lobo, An evolutionary theory for interpreting urban scaling laws. Cybergeo: Eur. J. Geogr. (2006).
49. G. Rushton, R. G. Golledge, W. A. Clark, Formulation and test of a normative model for the spatial allocation of grocery expenditures by a dispersed population. Ann. Assoc. Am. Geogr. 57, 389–400 (1967).
50. G. Rushton, Behavioral correlates of urban spatial structure. Econ. Geogr. 47, 49–58 (1971).
51. Kang, Y. et al. Multiscale dynamic human mobility flow dataset in the US during the COVID-19 epidemic. Sci. Data. 7, 390 (2020).
52. Bonnabel, S. Stochastic gradient descent on Riemannian manifolds. IEEE Transactions on Automatic Control 58, 2217–2229 (2013).
53. Adcock, A. B., Sullivan, B. D. & Mahoney, M. W. Tree-like structure in large social and information networks. in 2013 IEEE 13th international conference on data mining 1–10 (IEEE, 2013).
54. Christaller, W. Die Zentralen Orte in Süddeutschland: Eine Ökonomisch-Geogr. Unters. Über d. Gesetzmässigkeit d. Verbreitg u. Entwicklg d. Siedlgn Mit Städt. Funktionen. (Fischer, 1933).
55. Kiskowski, M. A., Hancock, J. F. & Kenworthy, A. K. On the use of Ripley's K-function and its derivatives to analyze domain size. Biophys. J. 97, 1095–1103 (2009).

*Supporting Information for*

# Intercity mobility reveals the hyperbolic geometry of city systems

Zhaoya Gong[1,2], Bin Liu[1,2,†], Chenglong Wang[1,2,†], Pengjun Zhao[1,2,*], Xiang Li[1,2], Kaixiang Zhang[1,2], Changcheng Kan[1,2], Xingjian Liu[3]

[1] School of Urban Planning and Design, Peking University Shenzhen Graduate School, Shenzhen, China
[2] Key Laboratory of Earth Surface System and Human-Earth Relations of Ministry of Natural Resources of China, Peking University Shenzhen Graduate School, Shenzhen, China
[3] Department of Urban Planning and Design, The University of Hong Kong, Hong Kong, China
† Contributed equally
* Corresponding author: Pengjun Zhao. Email: pengjun.zhao@pku.edu.cn

**This PDF file includes:**

Methods
Supplementary Note 1-2
Tables S1 to S3
Figures S1 to S17

## Methods

### Data and pre-processing

We constructed 14 intercity mobility networks for two countries, China and the United States, capturing both short-term and long-term human mobility patterns, with cities as the spatial unit for China and Metropolitan Statistical Areas (MSAs) for the United States. The datasets cover the pre-, mid-, and post-COVID19 pandemic periods, facilitating analysis of changes and recovery processes in urban systems reflected by human mobility under external shocks. For China, short-term mobility networks include three temporal datasets of inter-city movements (2019.10, 2020.10, and 2021.10), while long-term migration networks cover five periods (2018–2019, 2019–2020, 2020–2021, 2021–2022, 2022–2023). For the United States, short-term mobility networks consist of three MSA-level datasets (2019.3, 2020.3, and 2021.3), and long-term migration networks cover three periods (2019–2020, 2020–2021, 2021–2022). To ensure consistency in spatial scale, U.S. mobility data were aggregated to the MSA level to align with the city scale in China. The specific data sources and preprocessing steps are as follows:

The **short-term mobility dataset for China (D1)** is sourced from Baidu Map, a service provided by one of the largest internet companies in China. This dataset captures records of daily movements between cities by Baidu users during the last two weeks of October in 2019, 2020, and 2021. The data includes details such as origin, destination, date, and the number of movements. The real-time location information of users is derived from the geographic position associated with their requests. Baidu has enhanced the spatial accuracy of this data through data cleaning and algorithmic optimization. In terms of user sample representation, the dataset includes individuals of various genders (male, female, and unspecified) and all age groups, helping to mitigate biases in user sampling. For analysis, the data from these 14 days was averaged to obtain the mean intercity traffic during this two-week period.

The **long-term mobility dataset for China (D2)**, also sourced from Baidu Map, tracks annual intercity migration of China's working population from 2019 to 2023. It includes details such as origin city, destination city, and number of migrants, based on geographic positions from user requests. A migration event is identified when a user resides in a new city for an extended period (typically several months). This dataset is used directly to analyze long-term migration patterns.

The **short-term mobility dataset for the US (D3)** is sourced from published dataset in *Scientific Data* (https://github.com/GeoDS/COVID19USFlows).[51] It covers daily mobility patterns across counties from March 15 to April 15 in 2019, 2020, and 2021. The dataset includes details such as origin county, destination county, date, and number of movements, derived from anonymized mobile device location data. For analysis, daily movements over the 31-day period are averaged to estimate mean daily mobility. County-level flows are aggregated to the MSA level.

The **long-term mobility dataset for the US (D4)** is sourced from the U.S. Census Bureau (https://www.census.gov/data/tables/2020/demo/geographic-mobility/county-to-county-migration-2016-2020.html). It tracks annual migration between counties from 2019 to 2021. The dataset includes details such as origin county, destination county, and number of migrants, based on residence changes reported in Census surveys. A migration event is defined as a permanent relocation to a new county. For analysis, county-level migration flows are aggregated to the MSA level to examine long-term migration patterns.

## Hyperbolic geometric model of city systems

### *Poincaré ball*

Hyperbolic geometry, a non-Euclidean space characterized by constant negative curvature, offers a powerful mathematical foundation for modeling hierarchical systems. A key difference between hyperbolic and Euclidean geometry lies in how distances and areas scale with respect to the radius. In Euclidean space, the circumference and area of a disk grow linearly and quadratically, respectively, whereas in hyperbolic space, both increase exponentially with the radius. This exponential growth property makes hyperbolic space particularly well-suited for representing tree-like structures, where the number of nodes grows exponentially with distance from the root.

Among several common models of hyperbolic geometry, the Poincaré ball is especially advantageous for learning-based applications, as it provides a compact representation within the unit ball and supports smooth, gradient-based optimization.[35] Formally, the Poincaré ball defines an $n$-dimensional hyperbolic space within the unit ball:

$$\mathcal{B}^n = \{x \in \mathbb{R}^n : ||x|| < 1\} \tag{1}$$

where $||\cdot||$ denotes the Euclidean norm. The Poincaré distance between two cities $u, v \in \mathcal{B}^n$ is given by:

$$d(u,v) = arcosh\left(1 + 2\frac{||u-v||^2}{\left(1-||u||^2\right)+\left(1-||v||^2\right)}\right) \tag{2}$$

This distance between two cities $u, v$ with polar coordinates $(r_u, \theta_u)$ and $(r_v, \theta_v)$, can be approximated by the following equation when $r_v$ and $r_u$ are large:

$$d(u,v) \approx r_v + r_u + 2\, ln(\, sin(\frac{|\theta_u - \theta_v|}{2})) \tag{3}$$

Eq. 2 exhibits exponential sensitivity near the boundary of the Poincaré ball space. In such geometry, the root (highest order) of a hierarchy lies near the center and maintains short distances to all nodes, while low-order nodes are located progressively closer to the boundary. These geometric properties form the foundation upon which we build our hyperbolic embedding learning framework for intercity mobility networks.

### *Learning hyperbolic embeddings*

In this section, we propose an approach to embed the intercity network into a 2D Poincaré ball while preserving both hierarchical and proximity information. The intercity network is formulated as a weighted, directed graph $G = (V, E, W)$, where $V$ denotes the set of cities, $E$ represents the directed edges (i.e., connections between cities), and $W$ specifies the edge weights. Our objective is to learn a mapping function $f: V \to \mathcal{B}^2$ that embeds each city into the Poincaré ball such that the resulting configuration faithfully reflects the network's structural properties. The embedding process comprises three key components: (1) constructing weighted, directed random walks to capture representative transition sequences; (2) applying contrastive learning with domain-informed regularization to guide the embeddings; (3) utilizing Riemannian optimization methods tailored for hyperbolic geometry to ensure valid and stable updates. The following sections detail each step of the process.

**Weighted, directed random walks.** To extract meaningful patterns from the intercity network, we first generate a series of weighted, directed random walks that account for both edge asymmetry and connectivity strength. A weighted, directed random walk is defined as a sequence $s_1, s_2, \dots, s_i \in V$, where transitions are determined by a probability distribution. Given that simple random walks may become trapped in local regions, we choose a revisit-aware transition probability designed to balance local exploration and global connectivity.

Suppose a walk has just traversed edge $e_{u,v} \in E$ and is currently at node $v$ at step $i$. The transition probability to node $t$ at step $i+1$ is defined as:

$$P(s_{i+1} = t | s_i = v, s_{i-1} = u) = \frac{w_{v,t}^{\alpha}[\beta \mathrm{I}(t = u) + (1-\beta)\mathrm{I}(t \neq u)]}{\sum_{p \in N(v)} w_{v,p}^{\alpha}[\beta \mathrm{I}(p = u) + (1-\beta)\mathrm{I}(p \neq u)]}, \quad (4)$$

where $\mathrm{I}(\cdot)$ is a switch function, $N(v)$ denotes the neighbors of $v$, $\alpha$ controls the preference for high-weight edges, and $\beta$ modulates the tendency to revisit the previous node. To ensure sufficient coverage of the network, we initiate $T$ independent walks of length $L$ from each node.

**Contrastive learning with domain-informed regularization.** To preserve the structural information captured in the sampled walks, we adopt a contrastive learning framework with a soft ranking loss. Specifically, for each pair $(u, v)$ with high co-occurrence in the sampled sequences, we treat it as a positive sample and construct a corresponding negative sample set $D_u^-$, where $(u, v')$ pairs have low co-occurrence. The objective is to minimize the hyperbolic distances $d(u, v)$ of positive pairs while ensuring that the negative samples remain relatively farther apart. Formally, the contrastive loss is defined as:

$$L_c = -\sum_{(u,v) \in D^+} log(\frac{e^{-d(u,v)}}{e^{-d(u,v)} + \sum_{v' \in D_u^-} e^{-d(u,v')}}), \quad (5)$$

where $d(u, v)$ denotes the Poincaré distance between node embeddings in hyperbolic space, $D^+$ is the set of positive pairs, and $D_u^-$ denotes the corresponding set of negative samples for node $u$.

While this contrastive objective preserves the topological, directional, and intensity patterns in the data, it does not explicitly incorporate prior knowledge about the hierarchical nature of city systems. In real-world city systems, city size is not uniformly distributed: a small number of large cities dominate as global or national hubs, whereas a large number of small cities remain peripheral. We assume that city size has a lower bound, and thus the highest radial coordinate of nodes is bounded. It creates a necessary condition for the distribution of the hierarchy measure to converges to a power-law distribution even though it is not a sufficient one,[38]which facilitates the subsequential examination of city hierarchies' distribution, given that the commonly reported rank-size law of city systems.[17]

To encode these domain considerations, we introduce a regularization term that imposes a global constraint on node embeddings in hyperbolic space. Specifically, since the radial coordinate $r_v$ (Equation 2 in the main text) in the Poincaré ball corresponds to its latent hierarchy (i.e., smaller radius implies higher order):

$$L_R = \max_{v \in V}(r_v). \tag{6}$$

The final optimization objective integrates structural preservation with hierarchy regularization:

$$\min(L) = \min(L_c + \lambda L_R), \tag{7}$$

where $\lambda$ is a hyperparameter that controls the strength of the regularization. This unified formulation enables the learned embeddings to simultaneously capture the data-driven connectivity structure of the intercity network and the theory-informed hierarchy of city systems.

**Riemannian optimization in hyperbolic space.** Unlike Euclidean space, the Poincaré ball is a Riemannian manifold with constant negative curvature. Naively applying Euclidean gradient descent can lead to invalid updates that move points outside the Poincaré ball. To overcome this, we adopt Riemannian Stochastic Gradient Descent (RSGD),[52] which adapts learning with respect to the geometry of the underlying space. Specifically, RSGD performs updates in the tangent space of the manifold and maps them back to the Poincaré ball via exponential maps, which ensures all embeddings remain valid and allows the model to stably and efficiently learn representations that encode both global hierarchical structure and local spatial proximity.

***Model assessment***

**δ-hyperbolicity.** To assess the suitability of intercity networks for embedding in hyperbolic space, we begin by quantifying the tree-likeness of these networks using the well-established notion of δ-hyperbolicity.[53] This metric captures the extent to which a network exhibits an underlying hierarchical, tree-like structure, which is a critical prerequisite for meaningful representation in the Poincaré ball space. A lower δ-hyperbolicity value indicates that the network is closer to a tree and therefore more amenable to hyperbolic embedding.

Formally, the overall δ-hyperbolicity of a network is estimated by averaging δ-values computed over a large number of randomly sampled 4-tuples of nodes. For any given 4-tuple $(x, y, u, v)$, we first compute the following three combinations of pairwise shortest-path distances:

$$P = \{p_1, p_2, p_3\} = \{D(x,y) + D(u,v), D(x,u) + D(y,v), D(x,v) + D(y,u)\}. \tag{8}$$

These values are sorted such that $p_1 \leq p_2 \leq p_3$, and the δ-value for the 4-tuple is defined as:

$$\delta(x, y, u, v) = \frac{p_3 - p_2}{2}. \tag{9}$$

To obtain meaningful δ estimates, we first symmetrize the directed intercity network by averaging the in- and out-flows for each city pair. Then, we construct a flow-based distance matrix, where the distance ($D(u, v)$) between cities $u$ and $v$ is designed to reflect both their connectivity strength ( $w_{u,v}$ ) and degree centrality ( $W_u = \sum_k w_{u,k}$ , $W_v = \sum_k w_{v,k}$). Specifically, we define:

$$D(u,v) = \log\left(\frac{W_u \cdot W_v}{w_{u,v}}\right). \quad (10)$$

Finally, we estimate the overall δ-hyperbolicity of the network by averaging δ-values across a fixed set of randomly sampled node quadruples. This metric provides a meaningful way to evaluate whether the observed intercity network exhibits sufficient hierarchical structure to justify embedding in hyperbolic space.

**Reconstruction rate.** To evaluate whether the learned Poincaré embeddings faithfully preserve the structural properties of the original intercity network, we assess their reconstruction accuracy based on neighborhood consistency. For each city $v \in V$, we first identify its top-$N$ most strongly connected neighbors in the original intercity network, denoted as $S_{orig}(v)$. Similarly, we compute the top-$N$ nearest neighbors of $v$ in the embedding space according to Poincaré distances, denoted as $S_{emb}(v)$. To quantify the alignment between the two neighborhoods, we define the top-$N$ reconstruction rate, denoted as $R@N$, which measures the fraction of common neighbors:

$$R@N = \frac{\left|S_{orig}(v) \cap S_{emb}(v)\right|}{N}. \quad (11)$$

A higher value of $R@N$ indicates that the embedding space more accurately retains the local connectivity patterns observed in the original network. The final reconstruction performance is obtained by averaging $R@N$ across all cities in the network. This metric provides an intuitive and interpretable assessment of how well the learned hyperbolic representations preserve both proximity and structural dimensions in the intercity system.

**Tree structures**

***City-hinterland structures***

The city-hinterland relation is that a larger (high-order) city serves as the center providing high-order goods and services to a set of smaller (low-order) cities at different orders making up the hinterland, which may recursively form for smaller cities and their own hinterlands. Here, the orders are technically defined as a set of discrete levels of a hierarchy $\{\bar{r}_1, \bar{r}_2, \cdots, \bar{r}_R\}$, where $\bar{r}_1 > \bar{r}_2 > \cdots > \bar{r}_R$. The city-hinterland relation assumes the Hierarchy Property that larger cities provide all of the goods that smaller cities also provide. It means that a large city can be reviewed as a combination of itself and a series of smaller cities, each of which is at one lower order. Given that, a city-hinterland structure is defined as order dependent such that a central city links to low-order cities at only one certain order in the hinterland. In other words, a central city may have city-hinterland structures at multiple orders, collectively making up a nested hierarchy of cities (see Fig. S12a).

**City-hinterland relation.** For a central city $v$ at order $\bar{r}_i$, its hinterland is defined as a set of low-order cities $u$ served by $v$ by a relation link $\varepsilon(v,u)$:

$$\Omega(v, \bar{r}_i) = \left\{u \middle| \varepsilon(v,u), v \in \bar{r}_i, u \in \bar{r}_j, \bar{r}_i > \bar{r}_j\right\} \quad (12)$$

Here, $\Omega(v)$ encapsulates a set of low-order cities, which may be at different orders.

**City-hinterland structure.** The order-dependent city-hinterland structure is then defined by focusing on the linked hinterland cities at certain lower order. The city-hinterland structure of a central city $v$ at order $\bar{r}_j$ is given by:

$$\mathcal{H}(v,\bar{r}_j) = \{u|u \in \Omega(v,\bar{r}_i), u \in \bar{r}_j, \bar{r}_j < \bar{r}_i\} \Rightarrow \Omega(v,\bar{r}_i) = \bigcup_{j=i+1}^{R} \mathcal{H}(v,\bar{r}_j) \quad (13)$$

and thus the city–hinterland relation of $v$ can be obtained as the union of these city-hinterland structures.

### *Hierarchy measure and city order*

A city-size distribution is said to satisfy a power-law with exponent $\alpha$ if and only if for some positive constant $a$, the probability of a city size $C$ larger than $c$ is given by

$$\Pr(C > c) \approx ac^{-\alpha}, c \to \infty \quad (14)$$

If a given set of $n$ cities is postulated to satisfy such a power law, i.e., with city sizes distributed as in Eq. 14, and if these city sizes are ranked as $c_1 \geq c_2 \geq \cdots \geq c_n$, so that the rank $m_j$ of city $j$ is given by $m_j = j$, then it follows that a natural estimate of $\Pr(C > c_j)$ is given by the ratio, $j/n \equiv m_j/n$. So, by Eq. 14, we obtain the following approximation:

$$\frac{m_j}{n} \approx \Pr(C > c_j) \approx ac^{-\alpha} \Rightarrow \ln m_j \approx \beta - \alpha \ln c_j, \quad (15)$$

where $\beta = \ln an$.

According to Christaller, a central place system forms a hierarchy of central places (cities or towns), where places at lower order are hinterlands of places at higher order and hinterlands are exactly nested within one another according to a constant Nesting Factor $K$.[54] Given that there is one largest central place at the highest order in the central place system, the proportion of central places at order $\bar{r}_i$ is defined as:

$$PDF(\bar{r}_i) \approx \begin{cases} \frac{1}{n}, i = 1 \\ \frac{K^{i-2}(K-1)}{n}, i = 2,3,\cdots,R \end{cases} \quad (16)$$

where $K > 1$ is the nesting factor of hinterland. Then, the accumulative distribution $CDF(\bar{r}_i)$ defines the proportion of cities having higher order than $\bar{r}_i$:

$$\frac{i}{n} \approx CDF(\bar{r}_i) \approx \frac{K^{(i-1)}}{n} \Rightarrow \ln i \approx -\ln K + \ln K \cdot i \quad (17)$$

which can be naturally estimated by the ratio $\frac{i}{n}$, where $i$ is the rank of city at order $\bar{r}_i$.

Given that cities with larger size have high-order goods and services and thus are high-order cities, the city size rank is proportional to city order rank: $\frac{i}{n} \propto \frac{m_j}{n}$. By comparing Eq. 15 and Eq. 17 we obtain:

$$\ln c_j \propto \bar{r}_i \quad (18)$$

From Equation 2 in the main text, we have:

$$c_j \propto e^{-r_j} \Rightarrow -\ln c_j \propto r_j \quad (19)$$

Then, by comparing Eq.18 and Eq. 19 we obtain:

$$\bar{r_i} \propto -r_j \,, \tag{20}$$

which means the hierarchy measure and the city order are essentially equivalent. In other words, the hierarchy measure is a continuous representation of the city order.

Given the derived relationship between the hierarchy measure and the city order, we can discretize the hierarchy into city order by estimating $K$ (see Fig. S12b). The following steps are taken:

1) Compute the accumulated frequency $CDF(r)$ for each distinct radius $r$, counting the total number of cities with radii less than or equal to $r$.
2) Apply a logarithmic transformation to linearize the exponential relationship:

   $$logCDF(r) = (r-1)logK + \beta_0 \tag{21}$$

   where $\beta_0$ is a constant, representing the intercept of the linearized function.
3) Fit a linear model to the transformed data using linear regression, thereby estimating the slope $logK$.
4) Validate the estimated $K$ to ensure it satisfies the theoretical requirement $K > 1$. This constraint is crucial, as it confirms the expected structure of hierarchical urban systems, where cities at lower orders consistently outnumber those at higher orders.

   Based on the derived value of $K$, the continuous hierarchies are discretized into several intervals. A set of thresholds define the boundaries of these intervals, where the number of cities between adjacent intervals increases by a factor of $K$.

***Tree construction algorithm***

Next, we construct a tree structure that adheres to the following principles (Fig. S12c):

1) Each child city $u$, assigned to order $\bar{r_j}$, is connected to exactly one parent city $v$, which belongs to a higher order $\bar{r_i} > \bar{r_j}$, with the order difference not exceeding a threshold.
2) For each child city $u$ at order $\bar{r_j}$, the parent city is selected such that the Poincaré distance between them is minimized.

To further refine the hierarchy and optimize intercity connections, we introduce a boundary adjustment process aimed at minimizing the total Poincaré distance across all tree edges. This adjustment is conducted iteratively by examining each boundary between adjacent intervals. For a given boundary, we consider moving the last city of the preceding interval to the beginning of the following interval. If this adjustment reduces the total Poincaré distance of the tree, the move is accepted and the boundary is updated accordingly. Similarly, the first city of the following interval may be moved to the end of the preceding interval if doing so results in a shorter total distance. These adjustments are repeated until no further reduction can be achieved. The resulting tree connects all cities, with the root node representing the biggest central city at the highest order (Fig. S1a). This tree construction captures the intrinsic order dependent city-hinterland structures described.

***Nesting factor***

Given the definition of city-hinterland structure, we define its nesting factor. For the order-dependent city-hinterland structure $\mathcal{H}(v,\bar{r}_j)$ of center $v$ with respect to order $\bar{r}_j$ ,the nesting factor is defined:

$$NF = |\mathcal{H}(v,\bar{r}_j)| + 1 \tag{22}$$

where adding one considers the central city itself due to the Hierarchy Property assumption. Therefore, the nesting factor here is not globally constant but specific to each city-hinterland structure and order dependent.

***Hinterland range***

The range of hinterland for a city-hinterland structure delineates the boundary of the hinterland in terms of the proximity measure. It refers to the most distant proximity that a low-order city at certain order is willing to bear to reach the minimum level of mobility strength (the interaction threshold) to be admitted to the hinterland. Given the definition of city-hinterland structure, we can define the hinterland range as follows:

Given a central city $v$ and its order-dependent city-hinterland structure $\mathcal{H}(v,\bar{r}_j)$, for each low-order city $u \in \mathcal{H}(v,\bar{r}_j)$, the transformed proximity separation of the pair $(v,u)$ can be computed as $P_{v,u} = 2\ln\left(\sin\left(\frac{|\Delta\theta|}{2}\right)\right), \Delta\theta = \theta_u - \theta_v$, and so does its Poincaré distance $d(v,u)$. The range of a city-hinterland structure $\mathcal{H}(v,\bar{r}_j)$ is then determined by the pair $(v,u)$ that has the maximum proximity separation among a set of pairs whose Poincaré distances satisfy a threshold condition.

$$range(v,\bar{r}_j) = \max\{P_{v,u} | u \in \mathcal{H}(v,\bar{r}_j), d(v,u) \geq \gamma\} \tag{23}$$

where $\gamma$ denotes the threshold for Poincaré distance required for a city to be admitted into the hinterland.

**Analysis of hinterland ranges**

***Inferring a global K from hinterland ranges***

According to Christaller, the distances between central places of progressively higher orders increase by a factor of $\sqrt{K}$, where $K$ is the constant nesting factor for an ideal central place system.[54] The hinterland range is thus subject to this scaling factor. Ideally, the range of central places at order $\bar{r}_i$ is calculated as:

$$\Lambda(\bar{r}_i) = gK^{\frac{R-i}{2}} \Rightarrow \ln\Lambda(\bar{r}_i) = \ln g + \frac{R-i}{2}\ln K \tag{24}$$

where $g$ is the range at the lowest order $\bar{r}_R$ and $i = 1,2,\cdots,R$.

To infer $K$ from empirical hinterland ranges given our city-hinterland structure definition, we perform least-squares linear regression on the calculated ranges $range(v,\bar{r}_j)$:

$$range(v, \bar{r}_j) \propto a \cdot i + b, i = j + 1 \tag{25}$$

where $a = -\frac{1}{2}\ln K$ is the slope and $b = \ln g + \frac{R}{2}\ln K$ is the intercept. Then, we obtain:

$$K = e^{-2a} \tag{26}$$

***Characteristic ranges and geographic scales***

We define characteristic ranges as the specific intervals of hinterland ranges that exhibit significant concentration across multiple orders, corresponding to the peaks in their Kernel Density Estimation (KDE) (as shown in Fig. S5). We hypothesize that these characteristic ranges correspond to specific geographic scales, reflecting the spatial organization of urban systems. To validate this hypothesis, we first employ the cross-L function to test the statistical significance of these characteristic ranges. Subsequently, we verify the correspondence of these significant characteristic ranges with established geographic scales, such as regional boundaries.

**Cross L-function.** Unlike the idealized urban system by Christaller,[54] actual city–hinterland relations exhibit spatial nonstationary rather than equally spaced distribution. Low-order cities tend to cluster around higher-order central cities within certain proximity thresholds, giving rise to a series of statistically significant proximity intervals. The cross-L function assesses clustering or dispersion patterns of low-order cities around higher-order central cities using proximity metrics, enabling a comparison with characteristic ranges to confirm their statistical significance. We hypothesize that the preliminary characteristic ranges, are encompassed within these significant intervals. The process of cross L-function is as follows:

1) Formulation: Two point sets are defined: set $A$ comprises higher-order cities (order $\bar{r}_1, \bar{r}_2, \cdots, \bar{r}_{i-1}$) with $n_A$ cities, and set $B$ consists of cities at the lower order $\bar{r}_i$ with $n_B$ cities. The proximity between a higher-order city $a_i \in A$ (at angle $\theta_i$) and a lower-order city $b_j \in B$ (at angle $\theta_j$) is defined as $prox_{ij} = \min(|\theta_i - \theta_j|, 2\pi - |\theta_i - \theta_j|)$, ensuring the distance does not exceed $\pi$. All pairwise proximities of cities across set $A$ and $B$ construct an $n_A \times n_B$ symmetric matrix.
2) Cross L-function: The 1D cross L-function, defined in Eq. 27, is a normalized form of the 1D Cross K-Function in Eq. 28 (Fig. S7a,b). It ensures $L_{AB}(h) = 0$ under complete spatial randomness (CSR), with values greater than 0 indicating spatial clustering and values less than 0 suggesting a dispersed distribution.

$$L_{AB}(h) = \frac{K_{AB(h)}}{2} - h \tag{27}$$

$$K_{AB}(h) = \frac{1}{\lambda_B} \cdot \frac{1}{n_A} \sum_{i=1}^{n_A} \sum_{j=1}^{n_B} \mathcal{I}(prox_{ij} \leq h), \tag{28}$$

where $K_{AB}(h)$ represents the expected number of low-order cities located within a proximity distance $h$ of a randomly selected higher-order. $\lambda_B = \frac{\pi}{n_B}$ denotes the density of lower-order cities, with the total length of the circular domain measured as $\pi$. $\mathcal{J}(prox_{ij} \leq h)$ is an indicator function, equal to 1 if $prox_{ij} \leq h$ and 0 otherwise.

3) Significance testing via randomized distribution: To test significance, Monte Carlo simulations are conducted 999 times under the null hypothesis of CSR by generating random city configurations uniformly across the circular domain, with the number of simulated cities matching the size of set A and set B. For each simulation, $\hat{L}_{AB}(h)$ is calculated, and the observed $L_{AB}(h)$ is then compared against the simulated distribution to compute a p-value and determine whether the spatial pattern significantly deviates from randomness.
4) Validating the statistical significance of characteristic ranges: To identify statistically significant clustering intervals using the cross-L function, we first compute the derivation of $L_{AB}(h)$ and identify its local maxima[55] (where $L'_{AB}(h) = 0$), These intervals are deemed significant aggregated distribution if they satisfy two conditions: 1) $L'_{AB}(h) = 0$ and $L_{AB}(h) > 0$ indicates the boundary of aggregated distribution; 2) p-value < 0.001, indicating scales at which low-order cities exhibit significantly clustering around high-order cities. We then verify whether the proximities from preliminary characteristic ranges fall within these significant intervals to confirm their statistical significance.

**Identify corresponding characteristic geographical scales.** To investigate the correspondence between characteristic ranges and geographical scales, we observe that the geographical distances corresponding to these ranges exhibit concentrated distributions within specific intervals (Fig. S7c). Based on this observation, we first compute the cross-order proximity $prox_{ij}$ and the geographical distance $g_{ij}$ from the geographical coordinates of city pairs. The scatter plot is constructed with the geographical distance $g_{ij}$ on the x-axis and the proximity $prox_{ij}$ on the y-axis (Fig. S7c). For each significant characteristic range's corresponding proximity interval, we identify the peak of the geographical distance distribution within that interval as the characteristic geographical scale.

**Deriving isolines of interaction thresholds**

To justify the use of interpolation for generating isolines of interaction threshold in the $(r_v, range(v, \bar{r_j}))$ space, let $X = r_v$ and $Y = 2\,ln(sin(\frac{|\theta_u - \theta_v|}{2}))$, we can derive a linear relationship from Eq. 3:

$$Y = -X + [d(v,u) - r_u] \tag{29}$$

This implies a negative correlation between $X$ (the hierarchy of the central city) and $Y$ (the angular proximity) and suggests that isolines in the $(r_v, range(v, \bar{r_j}))$ space exhibit an approximately linear structure with negative slope.

Moreover, although $r_u$ (the hierarchy of hinterland city) varies across different orders, it tends to converge around a certain value within the same order. This phenomenon arises because each order is anchored by a series of central cities whose ideal hinterlands can be approximated as regions bounded by constant Poincaré distances:

$$r_v + r_u + range(v, \bar{r}_j) = d(v, u) \geq \gamma \tag{30}$$

Given that $r_v$ is constant and $d(v, u)$ is relatively fixed for a city-hinterland structure, when $range(v, \bar{r}_j)$ is maximized, $r_u$ the hierarchy of the low-order city delineating the hinterland range must be minimized. At the same time, $r_u$ has a lower bound as it must be greater than $r_v$ according to the city-hinterland relation. As a result, for city-hinterland structures at certain order the distribution of $r_u$ should be concentrated and relatively stable, and thus within a city order the variation of the intercept $d(v, u) - r_u$ in Eq. 29 depends more on the change of $d(v, u)$. This logic justifies the interpolation of isolines of interaction thresholds for city-hinterland structures by city order based on the distribution of $d(v, u)$. The empirical analyses show the concentrated distribution of $r_u$ reasonably supporting the above logic (see Fig. S16 and S17).

**Change Rate of Interaction Thresholds**

Let the corresponding boundary hinterland city $\hat{u} \in \bar{r}_j$ of center $v \in \bar{r}_i$ as the city that attains the maximum in Eq. 23:

$$\hat{u}(v, \bar{r}_i, \bar{r}_j) = \arg\max\{ P_{v,\hat{u}} \mid \hat{u} \in \mathcal{H}(v, \bar{r}_j), d(v, \hat{u}) \geq \gamma \} \tag{31}$$

The boundary city–hinterland pair is $(v, \hat{u}(v, \bar{r}_i, \bar{r}_j))$, and the Poincaré distance for this boundary pair is defined as: $d\left(v, \hat{u}(v, \bar{r}_i, \bar{r}_j)\right)$. For each central city at order $\bar{r}_i$, all boundary pairs are identified and their Poincaré distances calculated. We then summarize these Poincaré distances over all central cities $v \in \bar{r}_i$ by the median:

$$\tilde{d}(\bar{r}_i) = \text{median}\left\{ d\left(v, \hat{u}(v, \bar{r}_i, \bar{r}_j)\right) \mid v \in \bar{r}_i, \hat{u} \in \Omega(v, \bar{r}_i) \right\}, \tag{32}$$

where $\Omega(v, \bar{r}_i)$ is defined the same as in Eq. 12. We compute the discrete slope of $\tilde{d}(\bar{r}_i)$ between adjacent orders and the Change Rate of Interaction Threshold (CRIT) is defined as the mean of these adjacent-order slopes:

$$\text{CRIT} = \frac{1}{I} \sum_{i=1}^{I} \frac{\tilde{d}(\bar{r}_{i+1}) - \tilde{d}(\bar{r}_i)}{\bar{r}_{i+1} - \bar{r}_i} \tag{33}$$

## Supporting Text

### Supplementary Note 1: Model comparison

To assess the effectiveness of our proposed hyperbolic geometric model (HGM), we compare it with the classical Popularity-Similarity Optimization (PSO) model.[31] Model performance is measured using the top-N reconstruction rate (R@N), which captures the extent to which the nearest neighbors of each city in the embedding space align with those in the original network. Robustness is assessed across neighborhood sizes of N = 20, 30, 40, and 50 (Table S1).

In the CN intercity network (Table S2), the hyperbolic geometric model attains consistently higher R@N values in short-term data (2019–2021), reflecting a stronger ability to preserve local connectivity patterns. Our model's superiority extends to long-term data (2019–2023), particularly for larger neighborhood sizes. Overall, our hyperbolic geometric model provides more faithful preservation in the CN datasets.

In the US intercity network (Table S3), our model also has better performance than PSO does for short-term data. However, PSO achieves consistently higher R@N for long-term data. These results can be explained by the high δ-hyperbolicity measures for the US long-term data (Table S1), which indicate weak global hyperbolicity in the US long-term networks. The reduced hyperbolic property of this network limits the advantage of our hyperbolic geometric embeddings, although it still retains part of the hierarchical organization, as shown by its meaningful reconstruction accuracy.

Taken together, these findings provide strong evidence that the hyperbolic geometric model serves as an advantageous alternative to the classical PSO model for intercity network representation learning.

**Supplementary Note 2: Simulations of a theoretical hyperbolic geometric model**

Christaller's central place system has a homogeneous geography assumption, which leads to stationary structures of central place hierarchies with a uniform nesting factor and stationary ranges of hinterlands.[54] Here, we formulate this ideal model in terms of the hierarchy and range measures from our hyperbolic geometric model and simulate the stationary behavior in the trade-offs between the two measures.

In an ideal central place model with a given number of orders $R$, the range of a city at order $\bar{r}_v$ can be expressed as:

$$\Lambda(\bar{r}_v) = gK^{\frac{R-v}{2}} \tag{34}$$

where $g$ is the range at the lowest order, and $K$ is the constant nesting factor for the ideal central-place system. Assume that the hierarchy of the central city $v$ and the hinterland city $u$ are $r_v$ and $r_u$, respectively. Then, under the gravity model the interaction flow intensity between the two cities can be expressed as:

$$f \approx \frac{e^{-r_v}e^{-r_u}}{\Lambda^2} \tag{35}$$

To characterize the interaction in hyperbolic space, we introduce the Poincaré distance $d(v,u)$, which relates to the interaction flow intensity $f$ as:

$$e^{-d(v,u)} \propto f \tag{36}$$

Therefore, introducing a proportionality constant $C$ yields:

$$e^{-d(v,u)} = C\,\frac{e^{-r_v}e^{-r_u}}{\Lambda^2} \tag{37}$$

Taking the natural logarithm on both sides gives:

$$d(v,u) = r_v + r_u + (R-v)ln(K) - lnC + 2lng \tag{38}$$

If the hinterland city is regarded as located at order $\bar{r}_u$ , then the order of central city is $\bar{r}_v = \bar{r}_{u-1}$. Substituting into Eq. 38 yields:

$$d(v,u) = r_v + r_u + (R-u+1)ln(K) - lnC + 2lng \tag{39}$$

In Eq. 39, $C$ and $g$ are constants. Without losing generality, we may set $-lnC + 2lng = 0$, which gives the simplified form:

$$d(v,u) = r_v + r_u + (R-u+1)ln(K) \tag{40}$$

We employed the simulation approach to analyze the relationship between the hierarchy of central city and the hinterland range. Letting $X = r_v$, $Y = (R-u+1)ln(K)$, we examine the variations of X–Y curves under different parameter settings. Over all simulations, the model has a fixed number of orders $R = 7$. The hierarchy $r_v$ of the central city at the highest order $(\bar{r}_1)$ is assigned values in the range [0.5,7] with an interval 0.01. Assuming that the difference between the hierarchy measures of two adjacent orders is $\Delta$, the hierarchy of a hinterland city $r_u$ can be expressed as $r_u =$

$r_v + u \cdot \Delta$.

Fig. S13 shows that isolines of $d(v,u) = 10,11,12,13$ appear as straight lines with negative slopes, reflecting a linear relationship between city hierarchy and hinterland range. Fig. S14 further shows that as the nesting factor $K$ increases from 2.8 to 3.1, the slopes of the isolines decrease gradually, yet their linearity remains unchanged, indicating that $K$ influences the strength of the trade-off relationship between city hierarchy and hinterland range, but does not alter their inherent linear correlation. Fig. S15 illustrates that as the hierarchical difference $\Delta$ increases from 0.3 to 0.9, the absolute hierarchy differences between the central city and hinterland cities expand, while the slopes of the equal-distance lines remain constant.

The persistence of negatively sloped equal-distance lines across different parameter configurations confirms that the trade-off between hierarchy and range remains structurally stable in the hyperbolic geometric formulation. The nesting factor $K$ and hierarchy increment $\Delta$ only rescale the relationship but do not disrupt the underlying trade-off structures. This supports our model's capacity to reproduce the stationary behavior expected in the idealized central place system.

## Supporting Tables

**Table S1. Reconstruction rate of the HGM and normalized δ-hyperbolicity values for corresponding intercity networks.** The reconstruction rate evaluates how well the learned embeddings preserve the structural information of the original network. Values closer to 1 indicate better preservation of structural information; Normalized δ-hyperbolicity measures how well a network can be embedded in hyperbolic space. Lower δ-hyperbolicity value generally indicates a more tree-like or hierarchical structure.

**(a) Reconstruction performance and δ-hyperbolicity values for short-term intercity networks of China.**

| CN Short-term | | | | | |
|---|---|---|---|---|---|
| Year | Normalized δ-hyperbolicity | Neighborhood | | | |
| | | 20 | 30 | 40 | 50 |
| 2019 | 0.30 | 0.67 | 0.67 | 0.66 | 0.67 |
| 2020 | 0.26 | 0.66 | 0.65 | 0.65 | 0.65 |
| 2021 | 0.30 | 0.62 | 0.62 | 0.62 | 0.63 |

**(b) Reconstruction performance and δ-hyperbolicity values for long-term intercity networks of China.**

| CN Long-term | | | | | |
|---|---|---|---|---|---|
| Year | Normalized δ-hyperbolicity | Neighborhood | | | |
| | | 20 | 30 | 40 | 50 |
| 2019 | 0.27 | 0.43 | 0.45 | 0.47 | 0.48 |
| 2020 | 0.29 | 0.47 | 0.50 | 0.54 | 0.56 |
| 2021 | 0.31 | 0.49 | 0.52 | 0.56 | 0.59 |
| 2022 | 0.33 | 0.51 | 0.54 | 0.58 | 0.60 |
| 2023 | 0.30 | 0.46 | 0.50 | 0.55 | 0.58 |

**(c) Reconstruction performance and δ-hyperbolicity values for short-term intercity networks of USA.**

| US Short-term | | | | | |
|---|---|---|---|---|---|
| Year | Normalized δ-hyperbolicity | Neighborhood | | | |
| | | 20 | 30 | 40 | 50 |
| 2019 | 0.28 | 0.53 | 0.56 | 0.59 | 0.62 |
| 2020 | 0.29 | 0.53 | 0.57 | 0.61 | 0.63 |
| 2021 | 0.31 | 0.50 | 0.53 | 0.56 | 0.60 |

**(d) Reconstruction performance and δ-hyperbolicity values for long-term intercity networks of USA.**

| US Long-term | | | | | |
|---|---|---|---|---|---|
| Year | Normalized δ-hyperbolicity | Neighborhood | | | |
| | | 20 | 30 | 40 | 50 |
| 2019 | 0.37 | 0.27 | 0.31 | 0.31 | 0.30 |
| 2020 | 0.36 | 0.26 | 0.30 | 0.31 | 0.30 |
| 2021 | 0.34 | 0.30 | 0.32 | 0.32 | 0.32 |

**Table S2. Reconstruction rate of the HGM and the PSO models on CN datasets (2019–2023).** Reconstruction rate results for the CN intercity network are reported for short-term (2019–2021) and long-term (2019–2023) periods. The embedding performance is compared across neighborhood sizes N = 20, 30, 40, and 50. Higher values indicate better structural preservation.

| CN | | PSO | | | | HGM | | | |
|---|---|---|---|---|---|---|---|---|---|
| | | **20** | **30** | **40** | **50** | **20** | **30** | **40** | **50** |
| **Short-term** | **2019** | 0.51 | 0.57 | 0.59 | 0.61 | **0.67** | **0.67** | **0.66** | **0.67** |
| | 2020 | 0.54 | 0.58 | 0.59 | 0.63 | **0.66** | **0.65** | **0.65** | **0.65** |
| | 2021 | 0.52 | 0.56 | 0.59 | 0.61 | **0.62** | **0.62** | **0.62** | **0.63** |
| Long-term | 2019 | 0.39 | 0.46 | 0.50 | 0.54 | **0.49** | **0.50** | **0.58** | **0.59** |
| | 2020 | 0.41 | 0.48 | 0.51 | 0.55 | **0.47** | **0.50** | **0.54** | **0.56** |
| | 2021 | 0.46 | 0.50 | 0.55 | 0.56 | **0.49** | **0.52** | **0.56** | **0.59** |
| | 2022 | 0.47 | 0.50 | 0.53 | 0.56 | **0.51** | **0.54** | **0.58** | **0.60** |
| | 2023 | 0.45 | 0.49 | 0.53 | 0.55 | **0.46** | **0.50** | **0.55** | **0.58** |

**Table S3. Reconstruction rate of the HGM and the PSO models on the US datasets (2019–2021).** Reconstruction rates are computed from 2019 to 2021 for the US intercity network. The embedding performance is compared across neighborhood sizes N = 20, 30, 40, and 50. Higher values indicate better structural preservation.

| US | | PSO | | | | HGM | | | |
|---|---|---|---|---|---|---|---|---|---|
| | | **20** | **30** | **40** | **50** | **20** | **30** | **40** | **50** |
| Short-term | 2019 | 0.36 | 0.44 | 0.49 | 0.54 | **0.53** | **0.56** | **0.59** | **0.62** |
| | 2020 | 0.34 | 0.40 | 0.47 | 0.52 | **0.53** | **0.57** | **0.61** | **0.63** |
| | 2021 | 0.37 | 0.44 | 0.49 | 0.53 | **0.50** | **0.53** | **0.56** | **0.60** |
| Long-term | 2019 | **0.36** | **0.43** | **0.48** | **0.51** | 0.27 | 0.31 | 0.31 | 0.30 |
| | 2020 | **0.38** | **0.43** | **0.48** | **0.51** | 0.26 | 0.30 | 0.31 | 0.30 |
| | 2021 | **0.38** | **0.42** | **0.48** | **0.51** | 0.30 | 0.32 | 0.32 | 0.32 |

## Supporting Figures

**(a)**

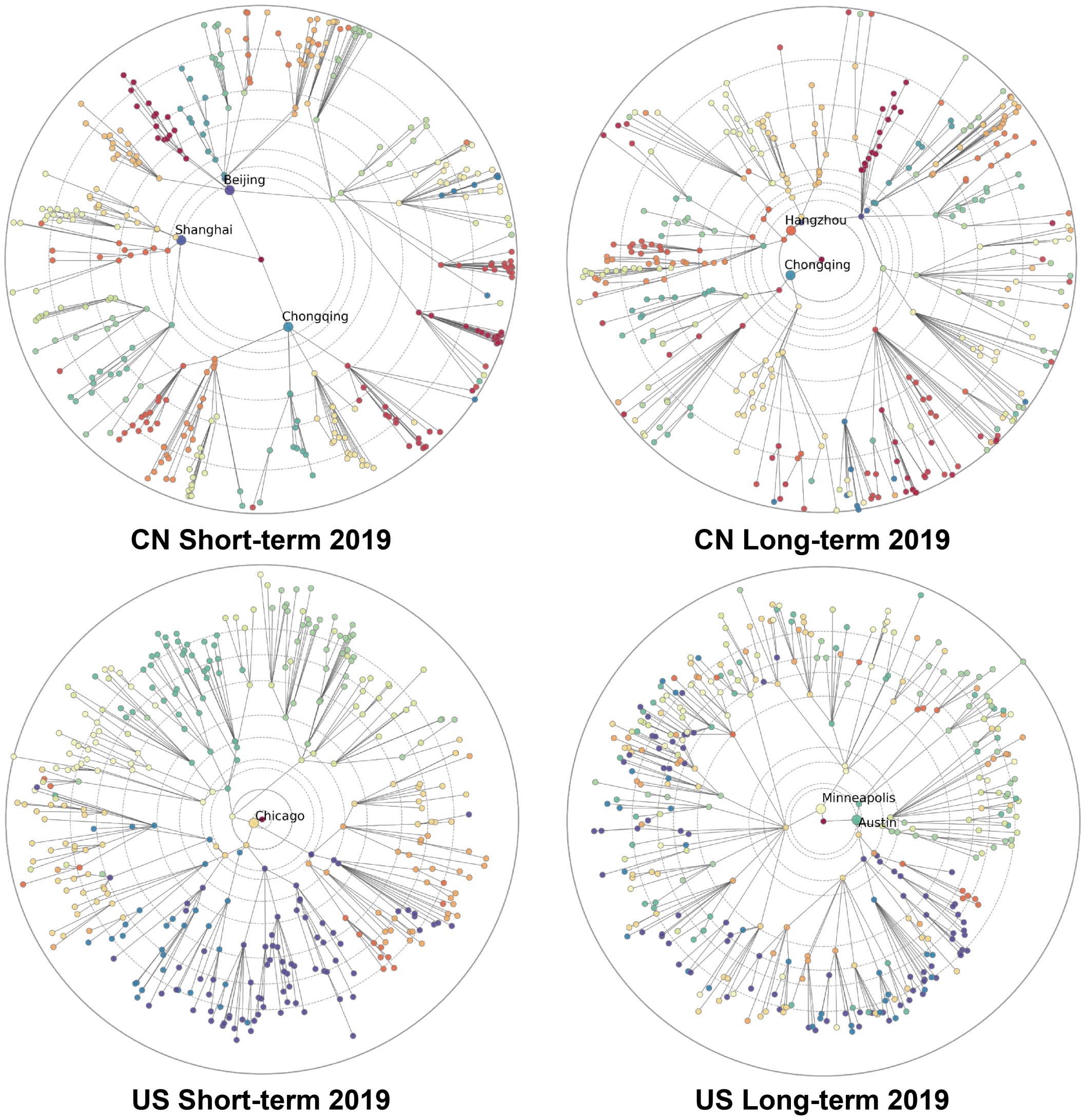

**(b)**

| Year | | CN | | | US | | |
|---|---|---|---|---|---|---|---|
| | | TF | VF | TF/VF | TF | VF | TF/VF |
| | 2019 | 26.88% | 80.95% | 33.21% | 16.68% | 83.66% | 19.93% |
| **Short-term** | 2020 | 27.52% | 81.03% | 33.97% | 16.55% | 83.63% | 19.79% |
| | 2021 | 27.22% | 80.06% | 33.99% | 16.09% | 85.38% | 18.84% |
| | 2019 | 10.36% | 82.66% | 12.54% | 13.33% | 83.11% | 16.04% |
| | 2020 | 11.30% | 84.70% | 13.14% | 9.50% | 80.39% | 11.82% |
| **Long-term** | 2021 | 11.69% | 85.71% | 13.63% | 15.34% | 85.46% | 17.95% |
| | 2022 | 13.30% | 85.39% | 15.57% | | | |
| | 2023 | 9.59% | 84.53% | 11.35% | | | |

**Fig. S1. Extracted tree structures and the proportion of mobility flows that they account for.** (**a**) Each node represents a city, with colors indicating different provinces or regions. Black dashed lines denote the edges connecting the nodes within the tree, while concentric red circles divide cities into hierarchical orders. (**b**) The Tree Flow (TF) is the proportion of mobility flows captured by the edges on the extracted tree structures for a mobility network; the Vertical Flow (VF) is the proportion of mobility

flows between cities at different orders, and the rest is the proportion of mobility flows between cities at the same orders.

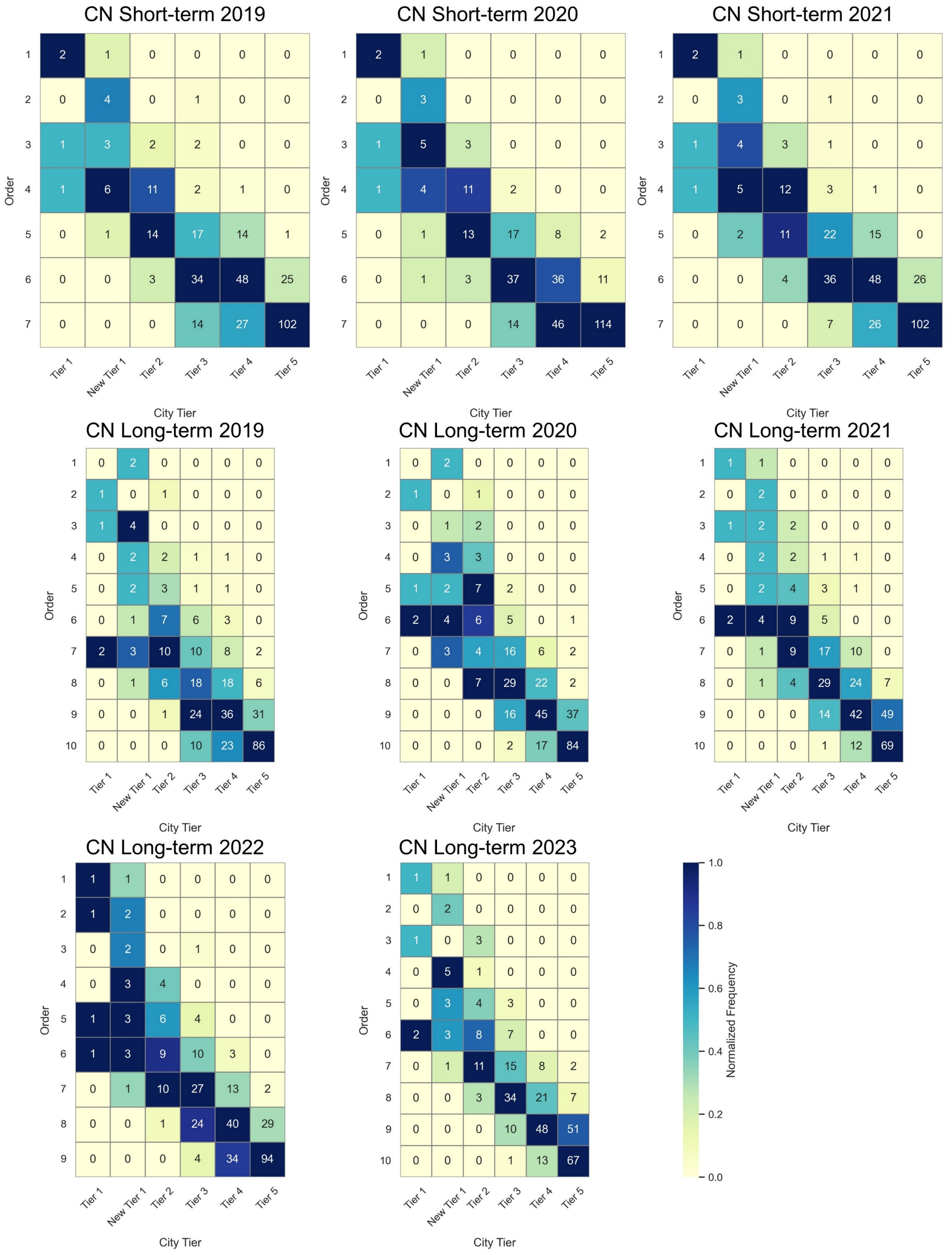

**Fig. S2. The association between city orders classified based on the hierarchy measure and the commonly accepted city ranking in practice.** Each cell of the heatmaps represents the number of cities that appear both in the city set of certain order in the row and in the city set of certain tier in the column. The color gradient represents

rescaled cell values normalized by the maximum per-tier count. Darker colors indicate a higher proportion within a tier. We observe clear correspondence between the derived city orders and the practical city ranking (sourced from https://www.datayicai.com/rankList).

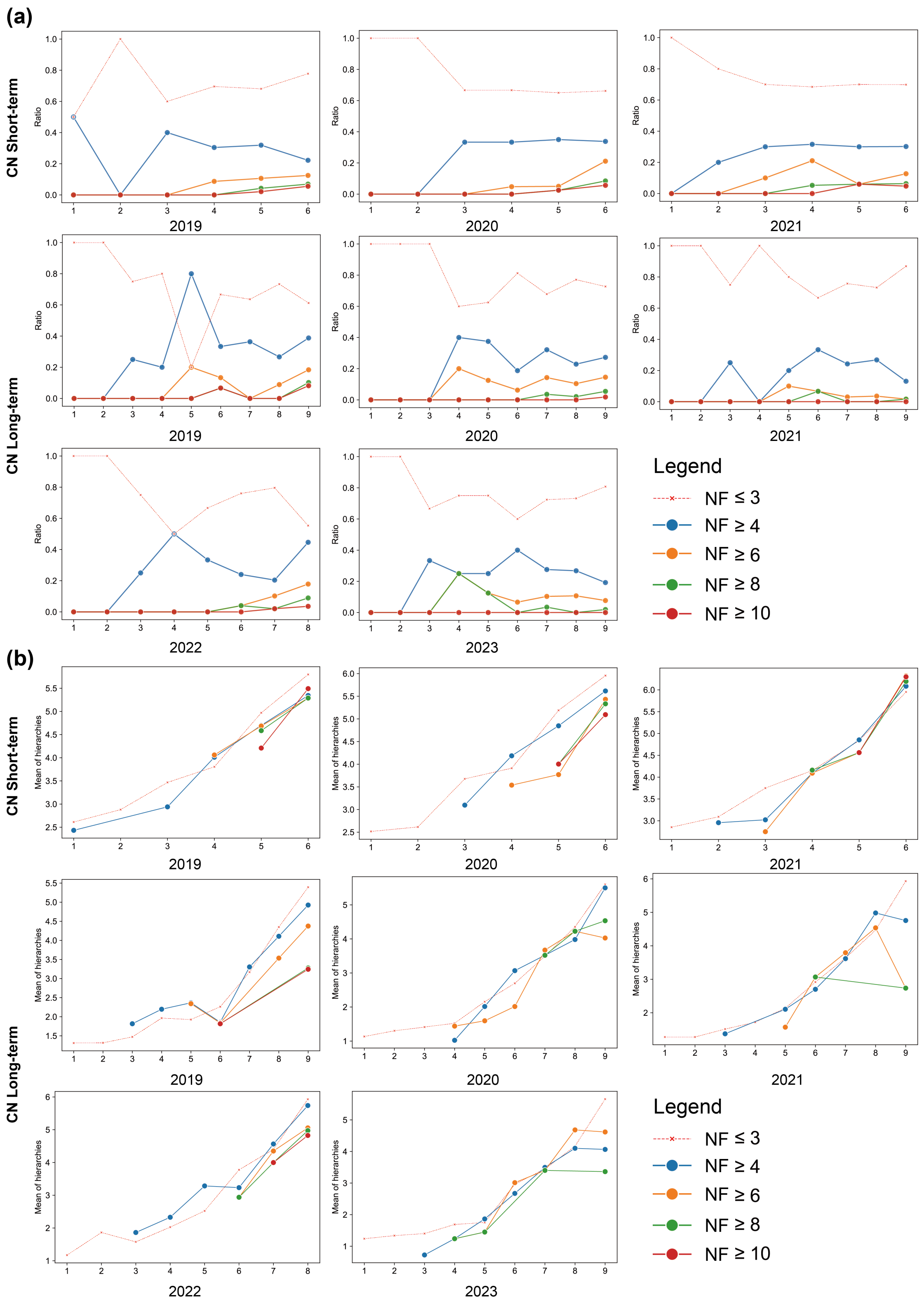

**Fig. S3. The variation of nesting factor (NF) the datasets of China.** (**a**) Proportion of city-hinterland structures with different NF values by city order (the horizontal axis). (**b**) The central city's average hierarchy for city-hinterland structures with different NF values by city order (the horizontal axis); a smaller value of hierarchy means a higher city order.

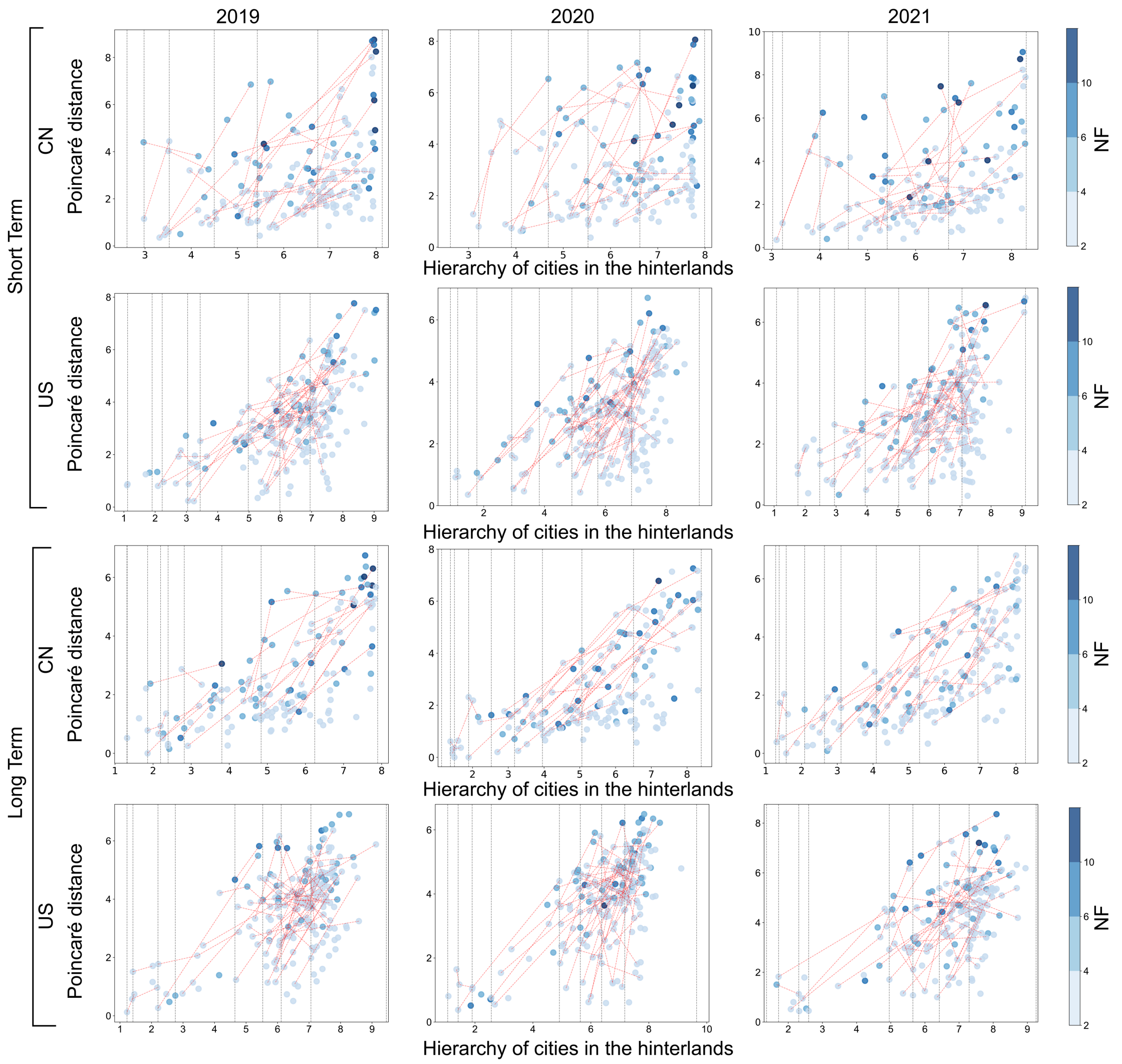

**Fig. S4. The interaction threshold (Poincaré distance on the vertical axis) of city-hinterland structures.** Each point represents a city-hinterland structure and its interaction threshold measured by Poincaré distance is the vertical coordinate; the hierarchy of the low-order city that defines the hinterland range is the horizontal coordinate. City-hinterland structures sharing the same central city are linked by a red dash line. The color gradient of points indicates the size of nesting factor (NF). Gray dash lines discretize the hierarchy measure into city orders. City-hinterland structures with large NF tend to be in low orders and associated with low level of interaction threshold (large Poincaré distance). In addition, the concentration of points in terms of the interaction threshold indicates that the level of interaction threshold decreases as the city order lowers. However, the decreasing patterns are different over years, between countries and types of mobility.

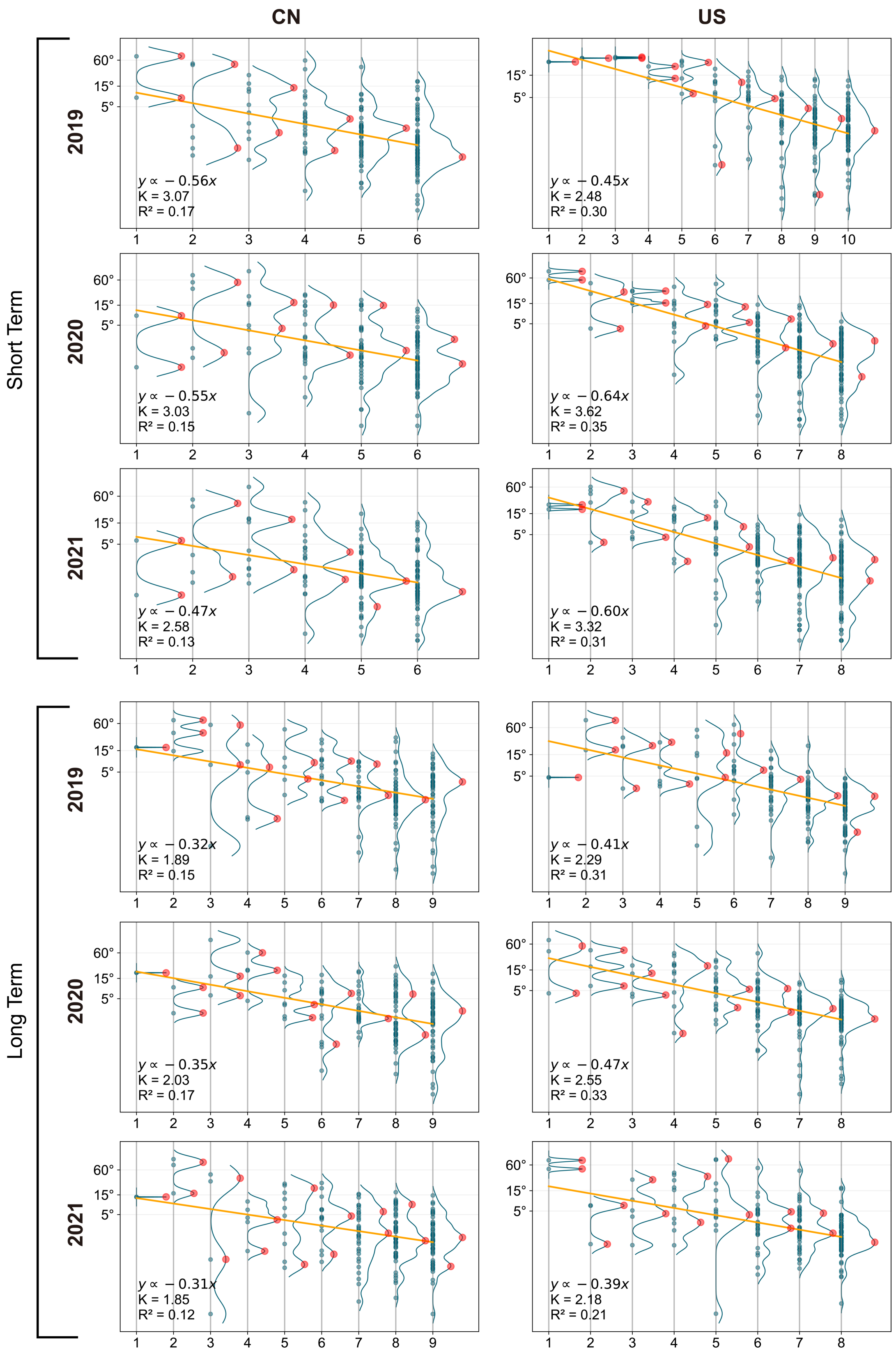

**Fig. S5. Distribution of ranges of city-hinterland structures by city order in city systems for all datasets.** Range is measured by proximity. Orange lines indicate linear modeling fittings with goodness of fit $R^2$ and global $K$ values are estimated based on the slope $k$ of fitted lines, i.e., $K = e^{-2k}$.

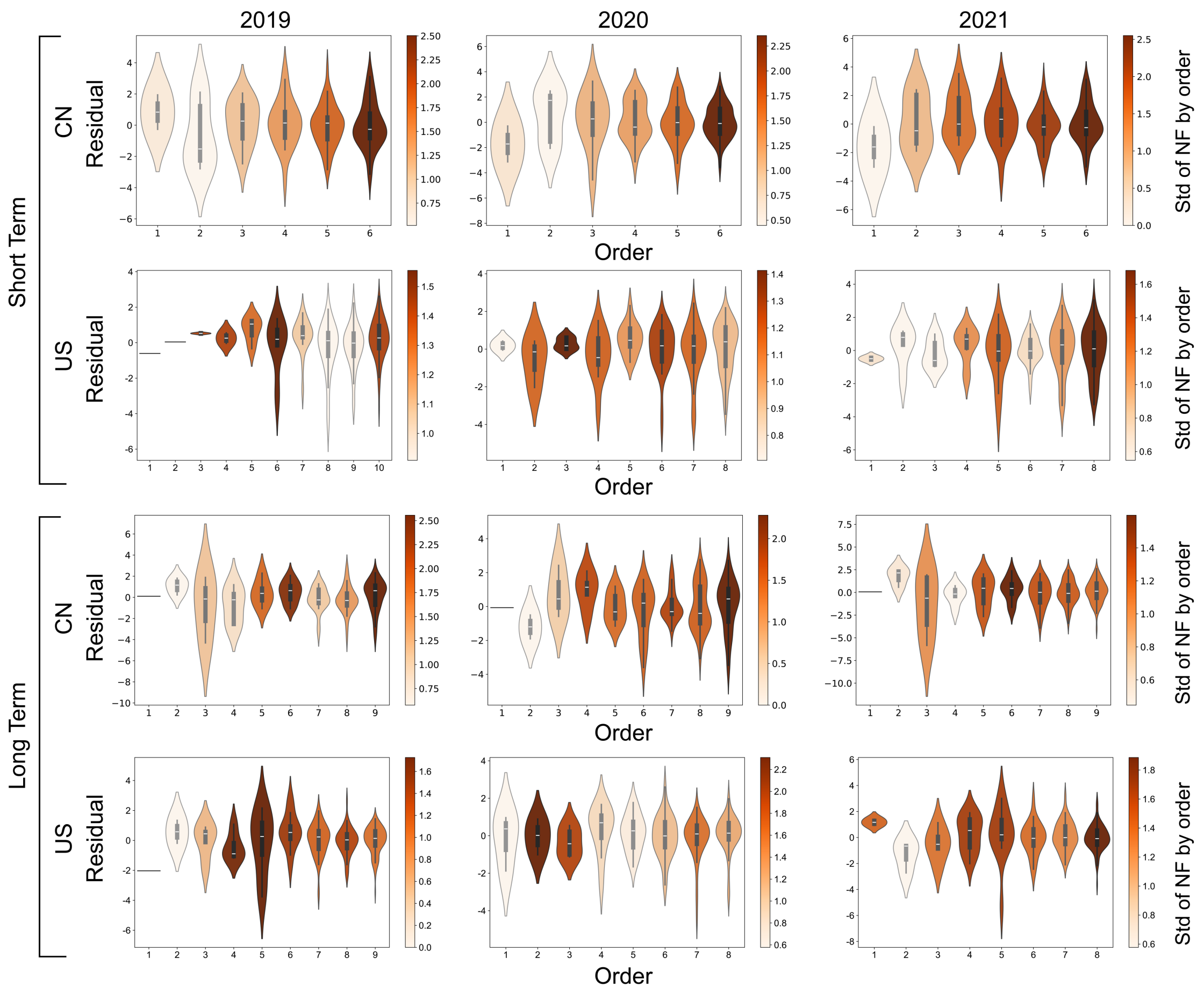

**Fig. S6. The nonstationary variations of residuals of fitted hinterland ranges by city order.** The residuals are from the linear model fittings for ranges of city-hinterland structures in Fig. S5. Standard deviation (std) of nesting factor values for city-hinterland structures by city order are indicated by color gradients.

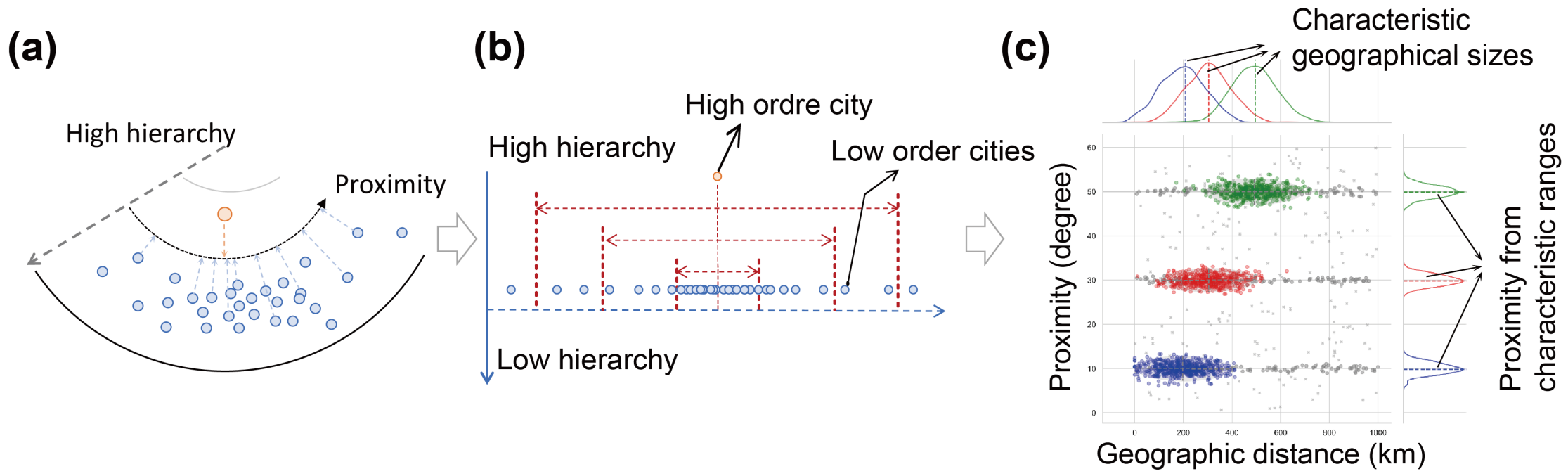

(d)

| Characteristic sizes | CN | | | US | | |
|---|---|---|---|---|---|---|
| | 2019 | 2020 | 2021 | 2019 | 2020 | 2021 |
| Metro-Area (proximity) | 2.81 ±0.35 | 3.55 ±0.63 | 3.58 ±0.70 | 4.42 ±1.13 | 2.44 ±0.73 | 4.28 ±1.147 |
| Size (km) | 263.16 ±50.90 | 312.77 ±66.37 | 350.41 ±99.18 | 293.37 ±112.06 | 211.14 ±65.07 | 278.21 ±97.20 |
| Urban-Region (proximity) | 14.13 ±1.01 | 12.94 ±1.12 | 15.26 ±1.13 | 22.72 ±2.08 | 17.16 ±0.81 | 19.79 ±2.80 |
| Size (km) | 527.93 ±131.84 | 530.53 ±121.31 | 660.43 ±261.95 | 602.77 ±198.80 | 692.83 ±203.20 | 733.79 ±207.49 |
| Division (proximity) | 61.56 ±1.31 | 63.16 ±1.743 | 55.18 ±1.57 | 54.27 ±2.39 | 58.37 ±1.04 | 53.69 ±0.55 |
| Size (km) | 899.95 ±131.71 | 1097.31 ±259.22 | 1008.50 ±174.32 | 1849.06 ±596.10 | 1661.68 ±361.39 | 2091.41 ±590.54 |

**Fig. S7. Identification of characteristic ranges and their corresponding geographical sizes (distance).** (**a~c**) Detecting clustering of low-order cities surrounding a high-order city within certain characteristic ranges with Cross L-function in a circular domain and identifying the correspondence between these ranges and characteristic geographical sizes. (**d**) Characteristic ranges (proximity) and their corresponding geographical sizes (distance). Each row lists a specific characteristic proximity alongside the type of geographical space it operates within. Columns distinguish China and the US. Comparing across rows reveals how similar types of proximity correspond to different spatial scales and structures under varying temporal and national contexts.

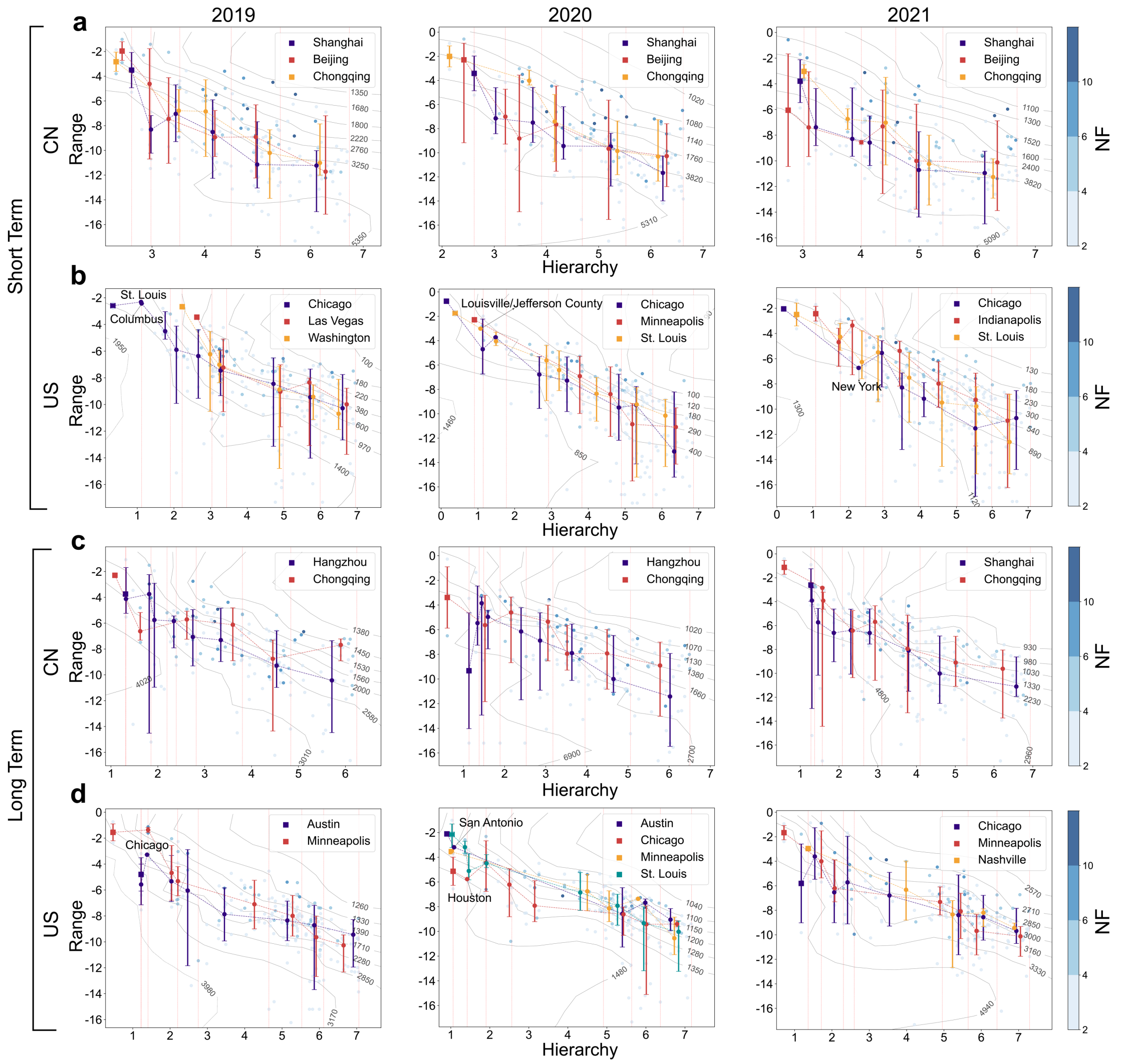

**Fig. S8. Trade-offs between city hierarchy and hinterland range.** Distributions of hinterland ranges for central cities by city order for different trees extracted for all datasets (**a, b, c, d**). Gray dash lines discretize the hierarchy measure into city orders. Trade-offs are indicated by large dots, whose coordinates are the median of hierarchy measures of central cities at each order versus the median of their hinterland ranges. Error bars represent the 10th–90th percentile of the distribution of ranges. Square dots are the root cities (city names labeled in legend) for extracted trees. Dots are linked by dashed lines for each tree. If there is only one central city at an order, its name is annotated. Gray lines represent derived isolines of interaction thresholds with annotations indicating the estimated mobility strength, which facilitate temporal comparisons. Over time, the interaction strengths of the entire city systems for both countries were significantly reduced due to the pandemic in 2020 and recovered (not to the level in 2019) in 2021. Isolines are derived based on the interaction thresholds of all city-hinterland structures represented by small dots in the background with color gradients indicating their nesting factor (NF) values.

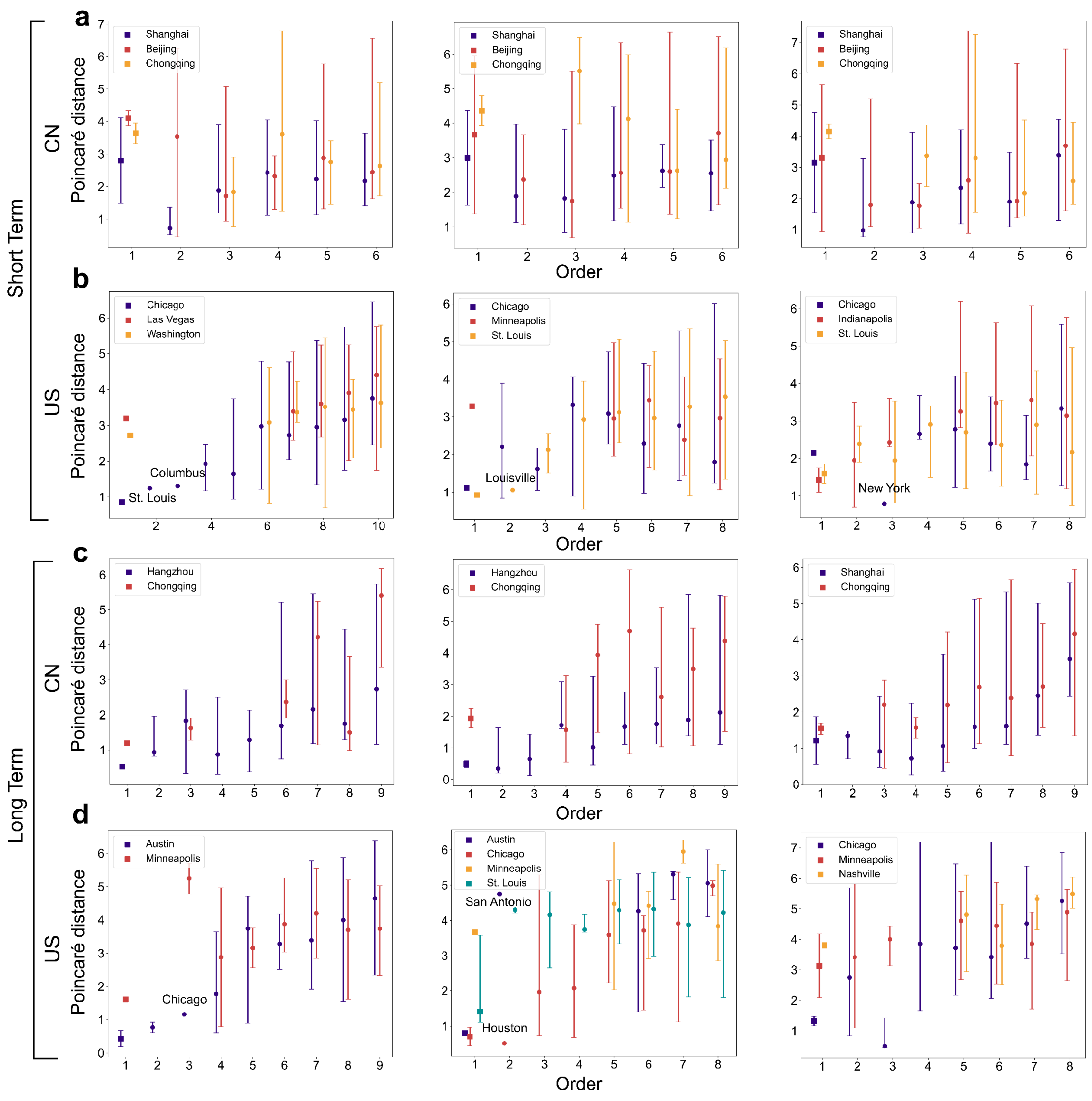

**Fig. S9. The change of interaction thresholds by city order.** Distributions of interaction thresholds of city-hinterland structures by extracted trees and by city order for all datasets (**a, b, c, d**). Interaction threshold is measured by Poincaré distance. A dot indicates the median of interaction thresholds of city-hinterland structures whose central cities at the same order and its error bars represent the 10th–90th percentile of the distribution of interaction thresholds. Square dots are the root cities (city names labeled in legend) for extracted trees, which are color-coded. If there is only one central city at an order, its name is annotated.

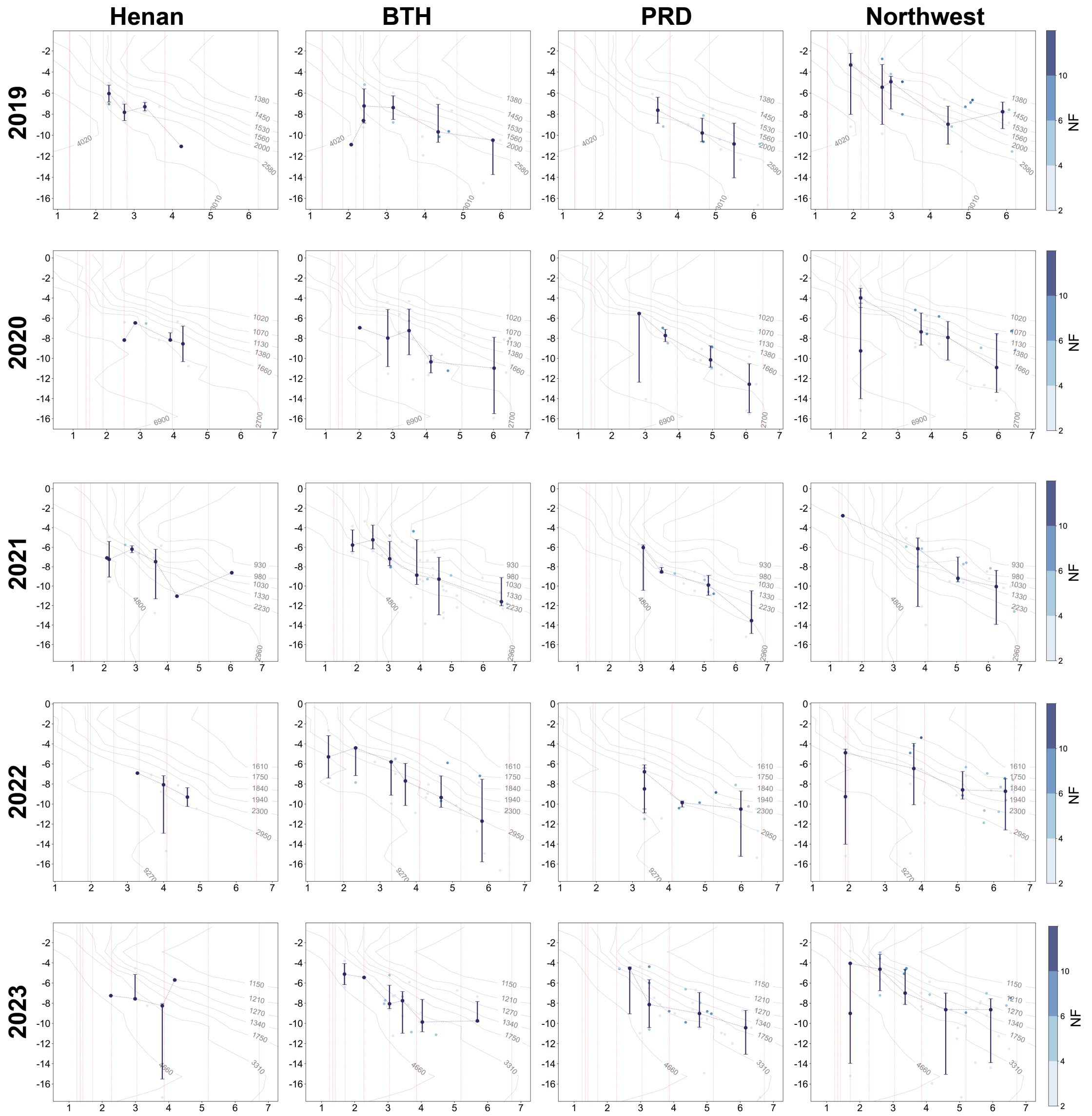

**Fig. S10. Trade-offs between city hierarchy and hinterland range in different regions of China.** This figure presents a regional comparison of the trade-offs between city hierarchy and hinterland range for major regions in China for long-term datasets. Each column corresponds to a specific region: Henan region, Beijing-Tianjin-Hebei Region (BTH Region), Pearl River Delta Region (PRD Region), and Northwest China. The analysis follows the same procedure as in Fig. S8. For each specific region, the $(r_v, range(v, \bar{r}_j))$ space is constructed based on the central city $v$ belonging to that region. Scatter points indicate the median hierarchy and range at each order, with error bars representing the 10th–90th percentile range. Isolines (gray) for the corresponding datasets from Fig. S8 are presented for reference.

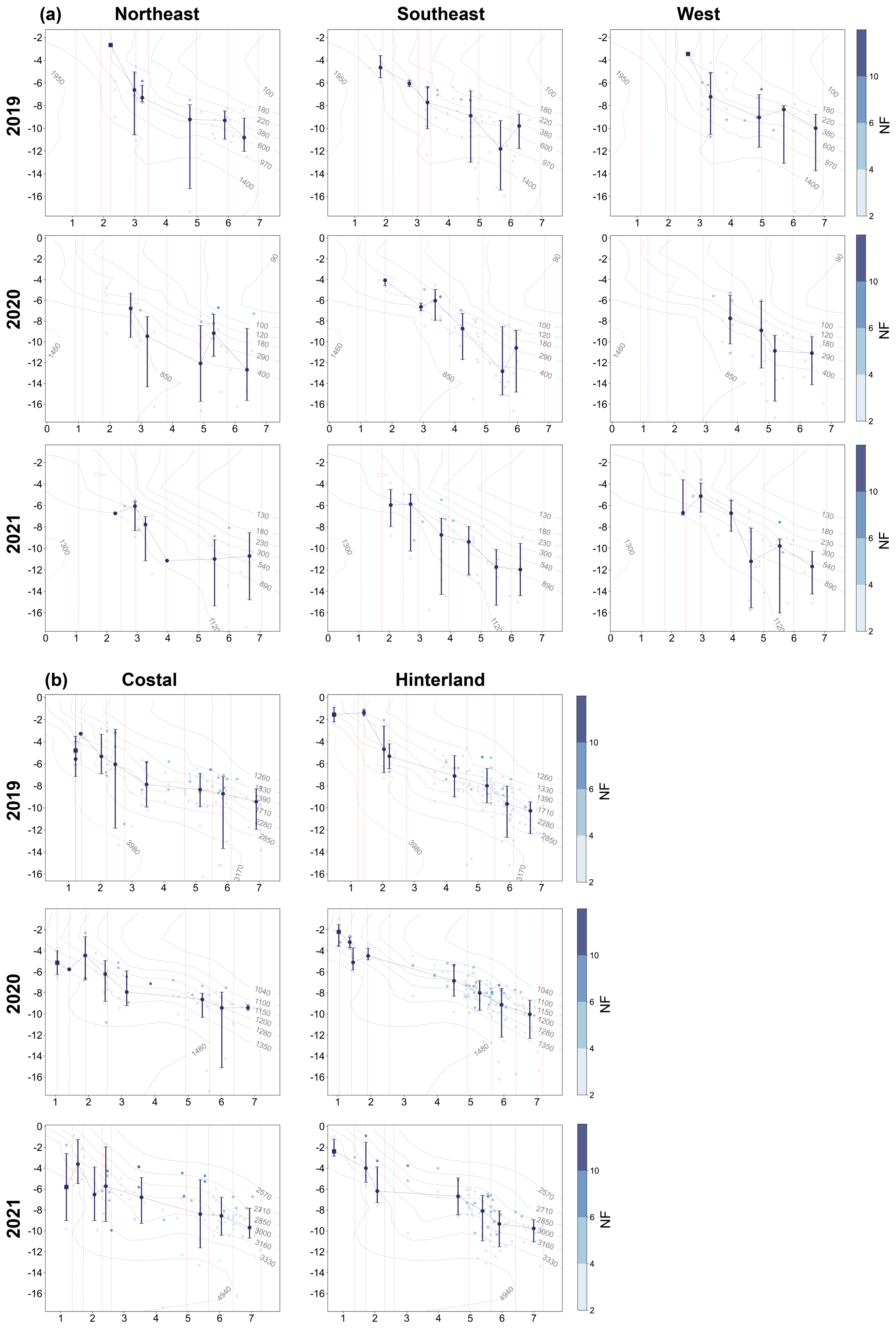

**Fig. S11. Trade-offs between city hierarchy and hinterland range in different regions of the US.** (**a**) Results of Northeast, Southeast, and West regions for short-term datasets; (**b**) Results for the coastal and hinterland regions for long-term datasets.

Isolines (gray) from Fig. S8 are presented for reference.

**(a) Illustration of the city–hinterland relation and the order-dependent city–hinterland structure**

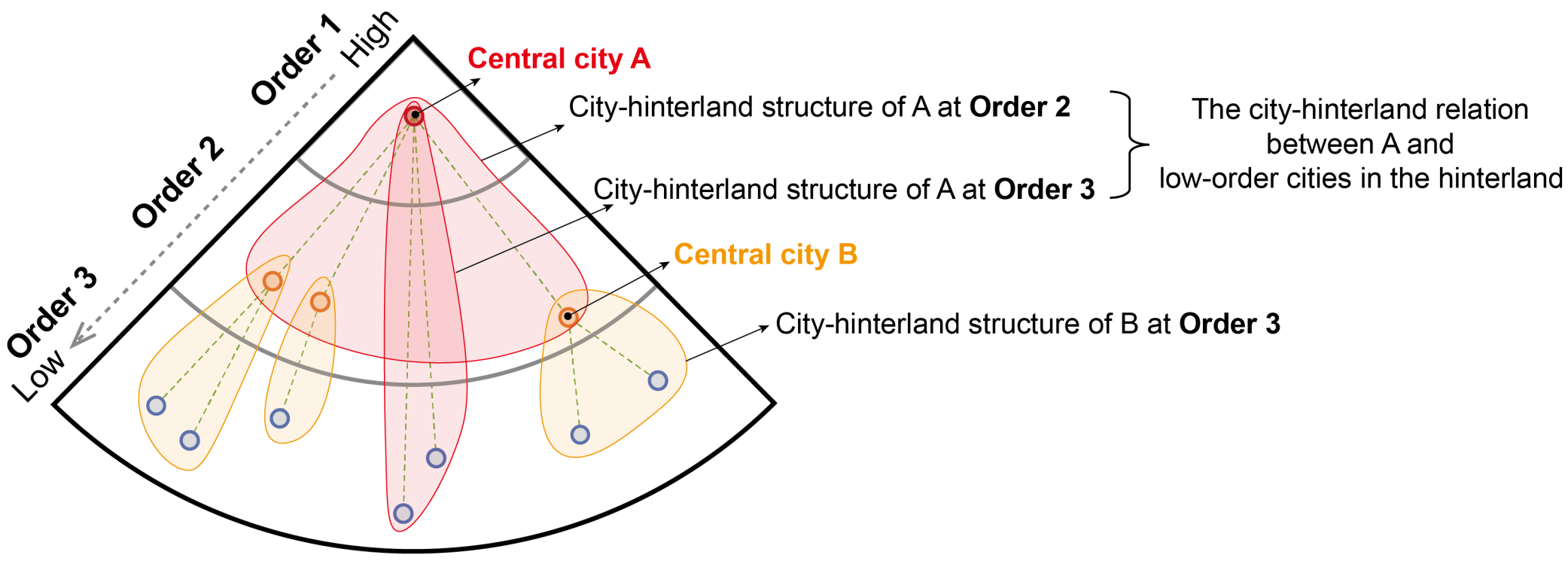

**(b) Discretizing hierarchy into order**

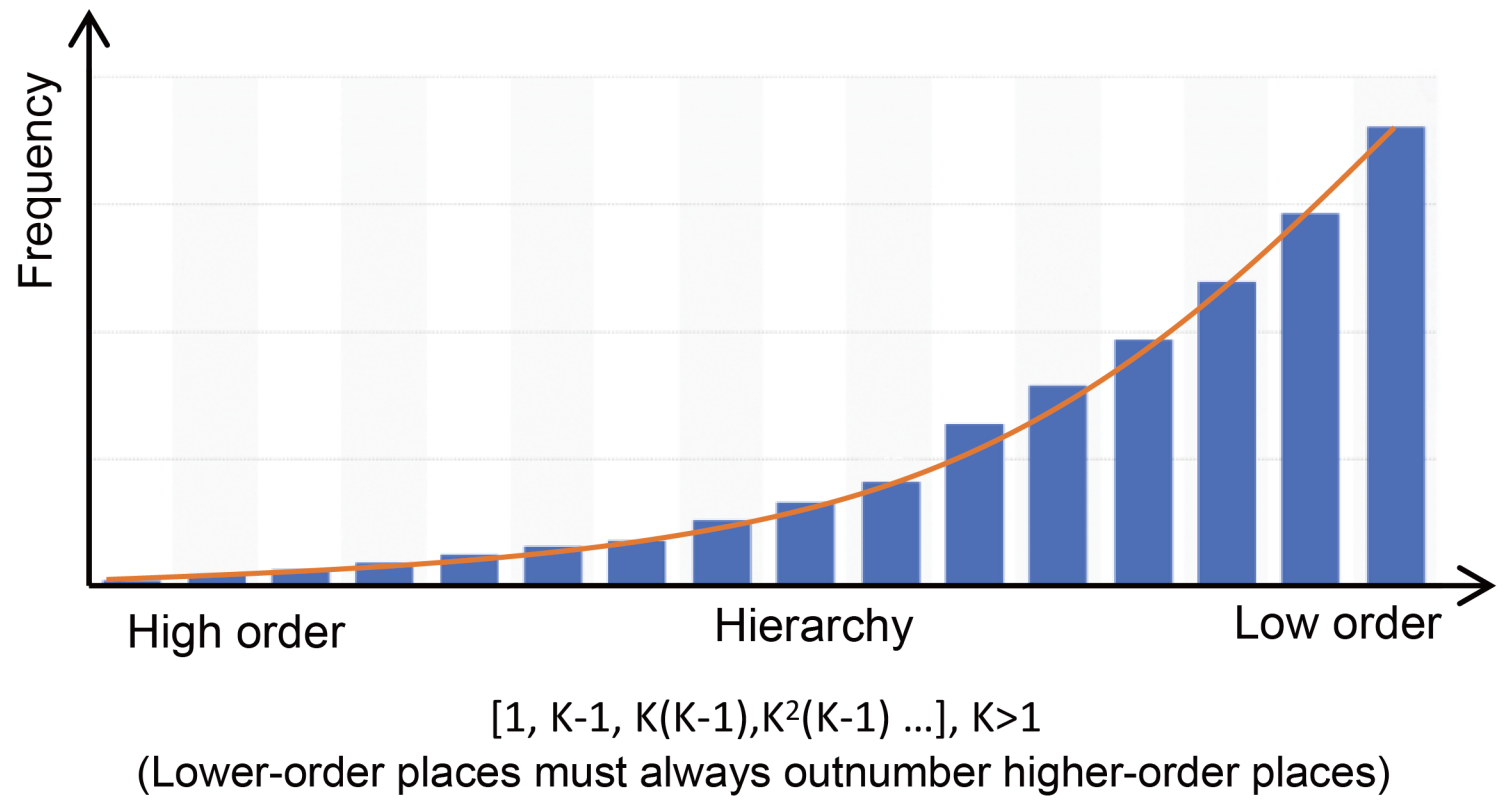

**(c) Algorithmic procedures of tree construction for city hierarchies**

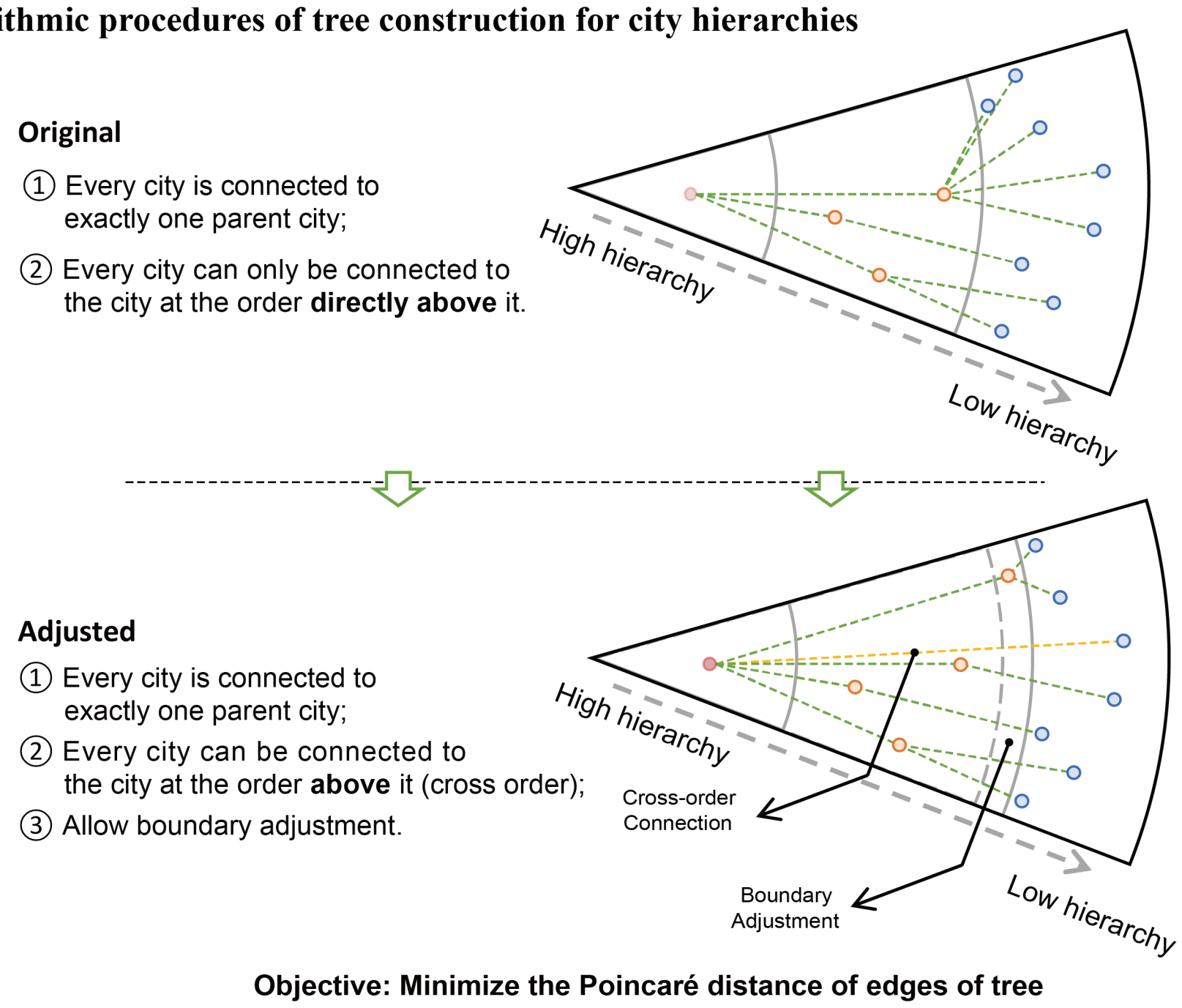

**Fig. S12. Extracting tree structures of city-hinterland relations**

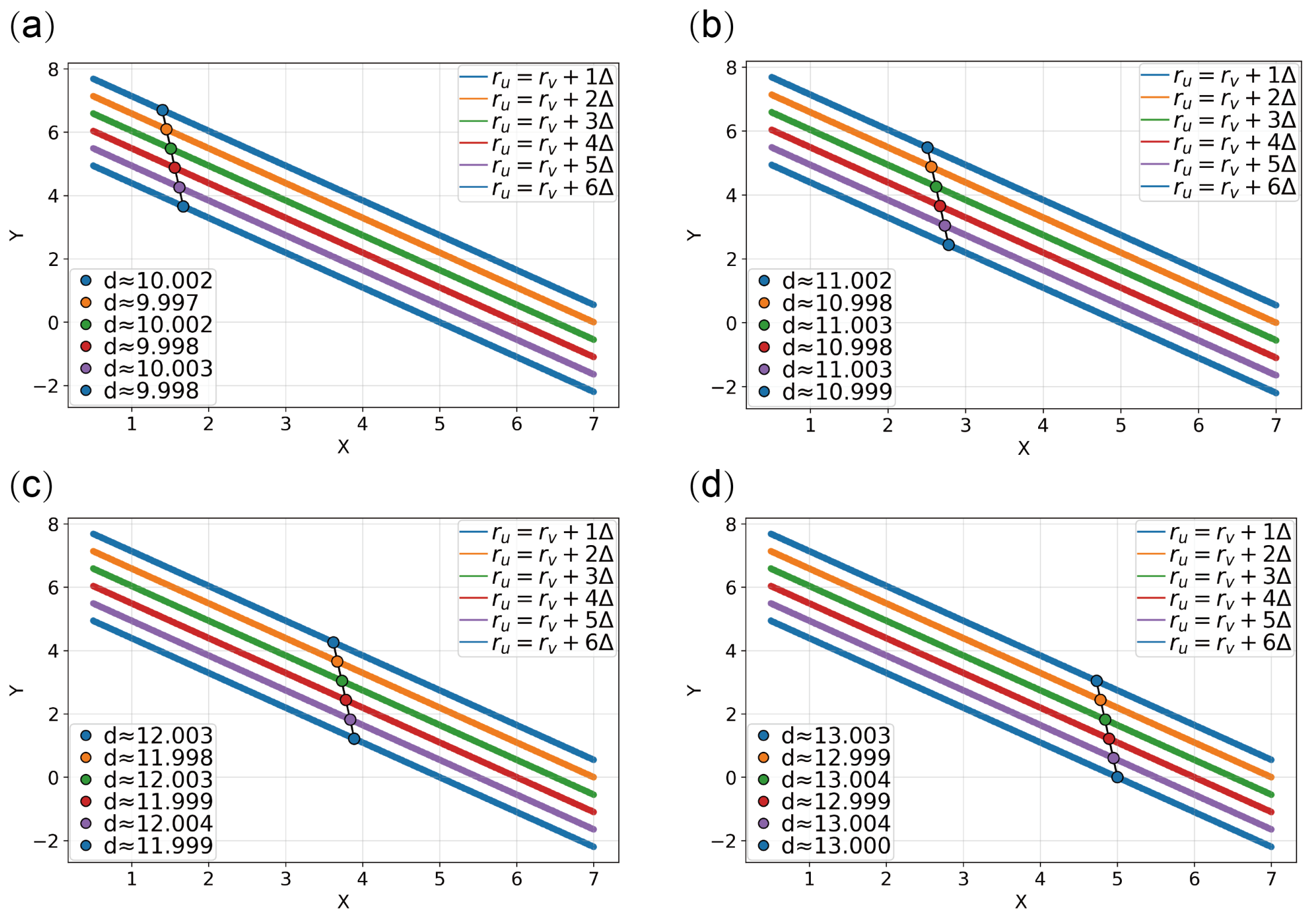

**Fig. S13. Isolines of $d(v,u)$ with $K$=3 and $\Delta$=0.5: (a) $d$=10, (b) $d$=11, (c) $d$=12, and (d) $d$=13**

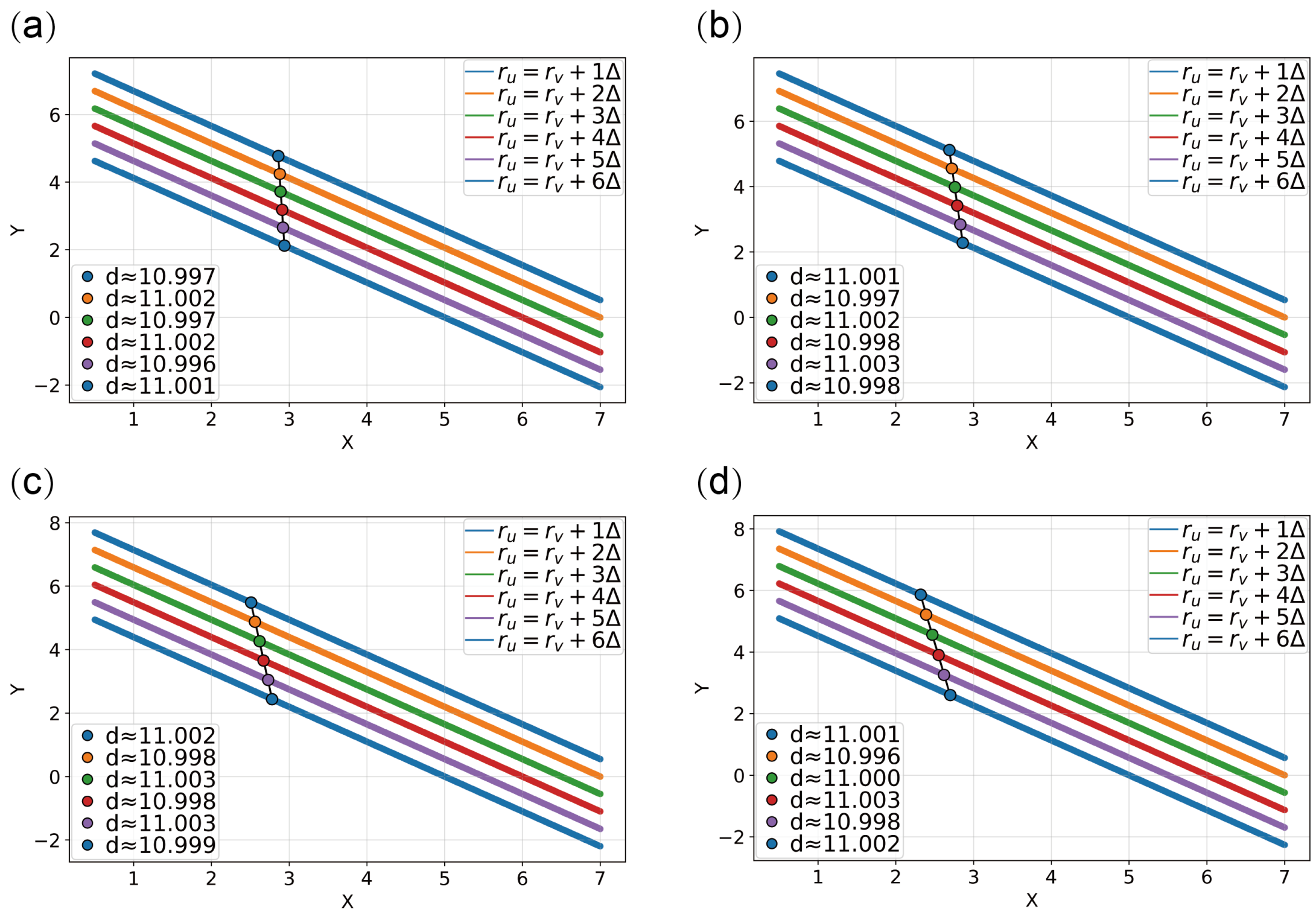

**Fig. S14. Isolines of $d(v,u)$=11 with $\Delta$=0.5 and varying $K$: (a) $K$=2.8, (b) $K$=2.9, (c) $K$=3, and (d) $K$=3.1**

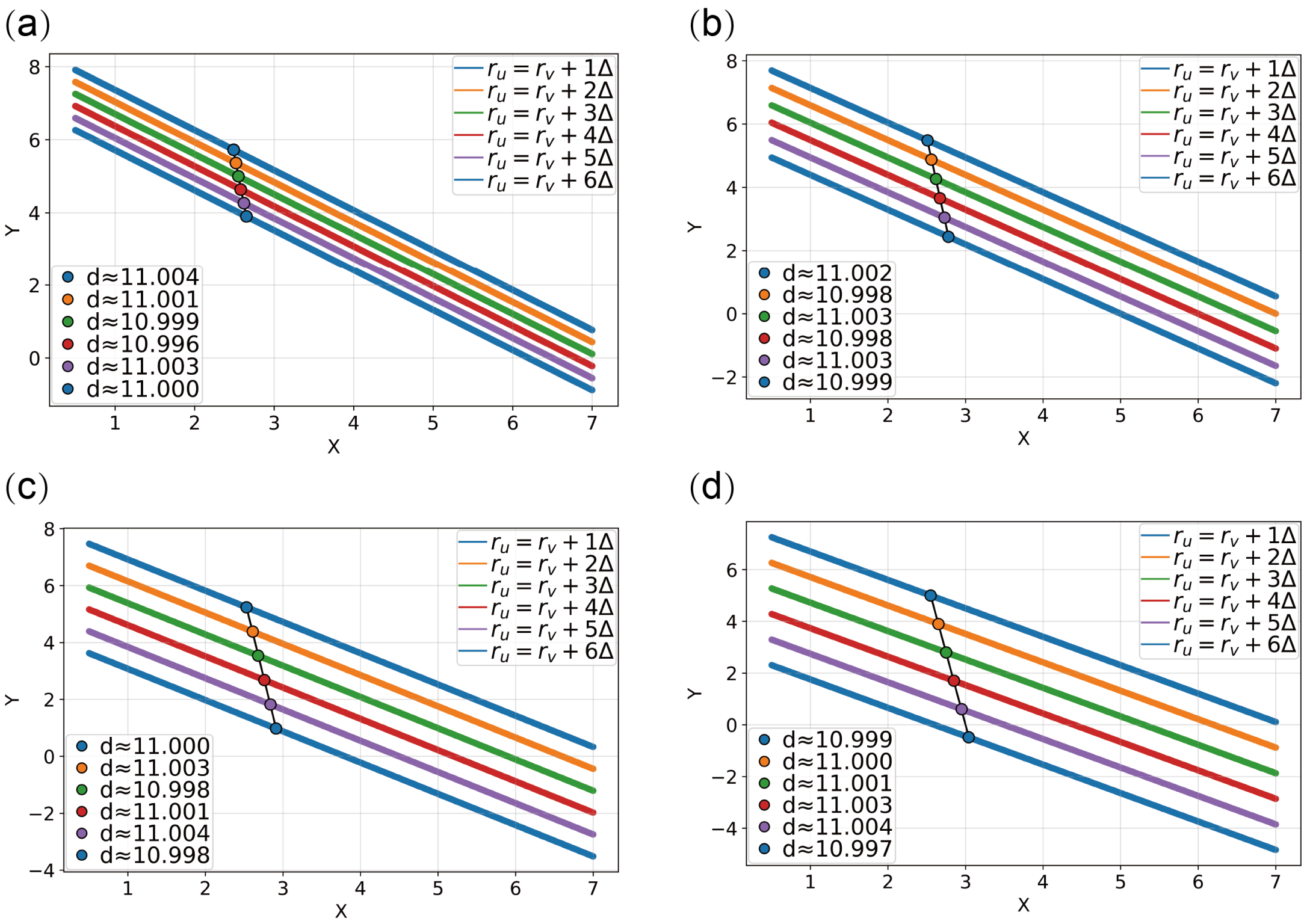

**Fig. S15. Isolines of $d(v,u)$=11 with $K$=3 and varying $\Delta$: (a) $\Delta$=0.3, (b) $\Delta$=0.5, (c) $\Delta$=0.7, and (d) $\Delta$=0.9**

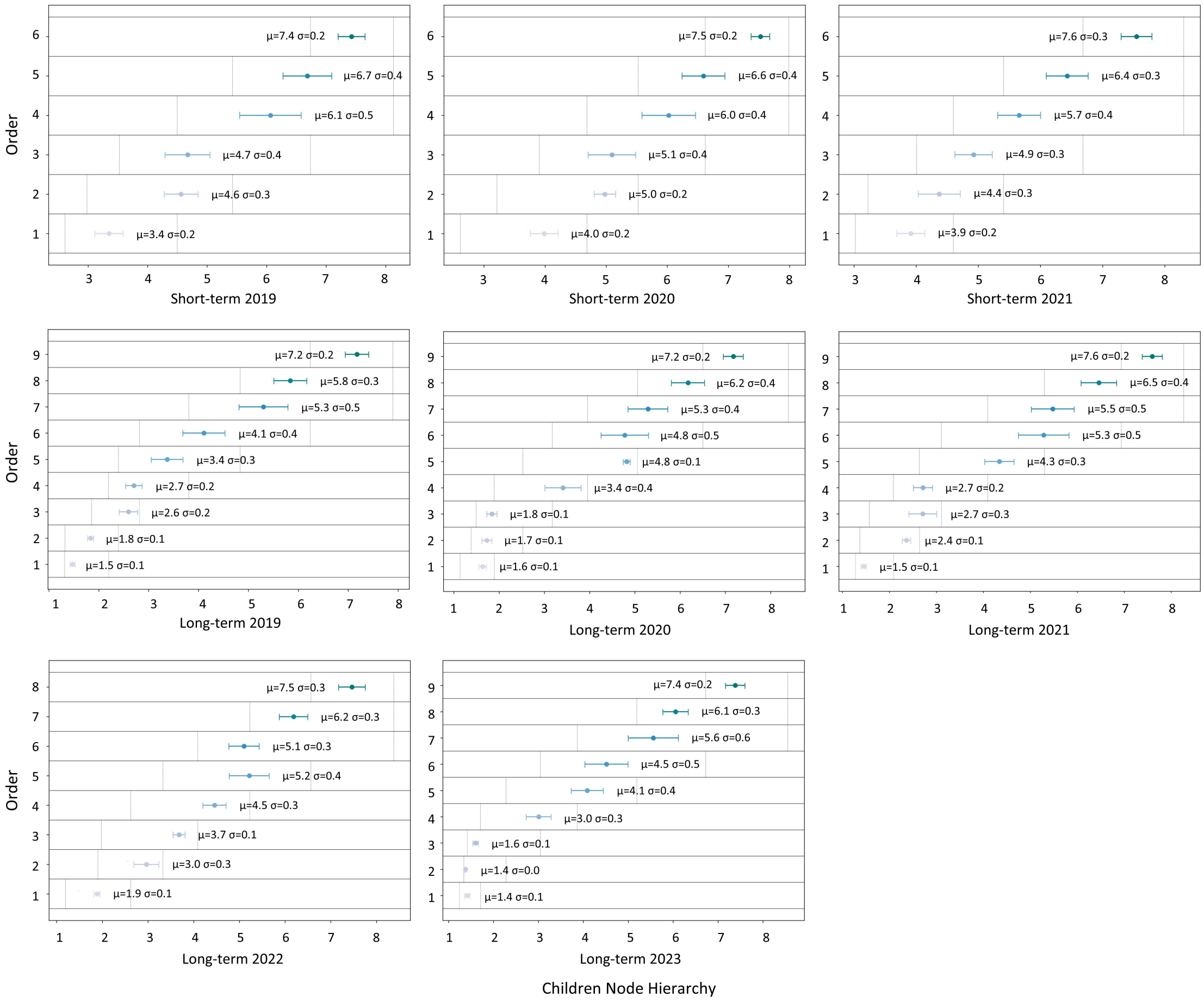

**Fig. S16. Distributions of hierarchy measures of the hinterland city that defines the range by city order for long-term and short-term datasets of China.** Each subplot shows results from a dataset. For each order, the mean (μ) and standard deviation (σ) of the hierarchy measures are displayed. Due to the maximum limit across city orders, the vertical gray dashed lines represent the ideal boundaries for the hierarchy distribution of hinterland city that defines the range.

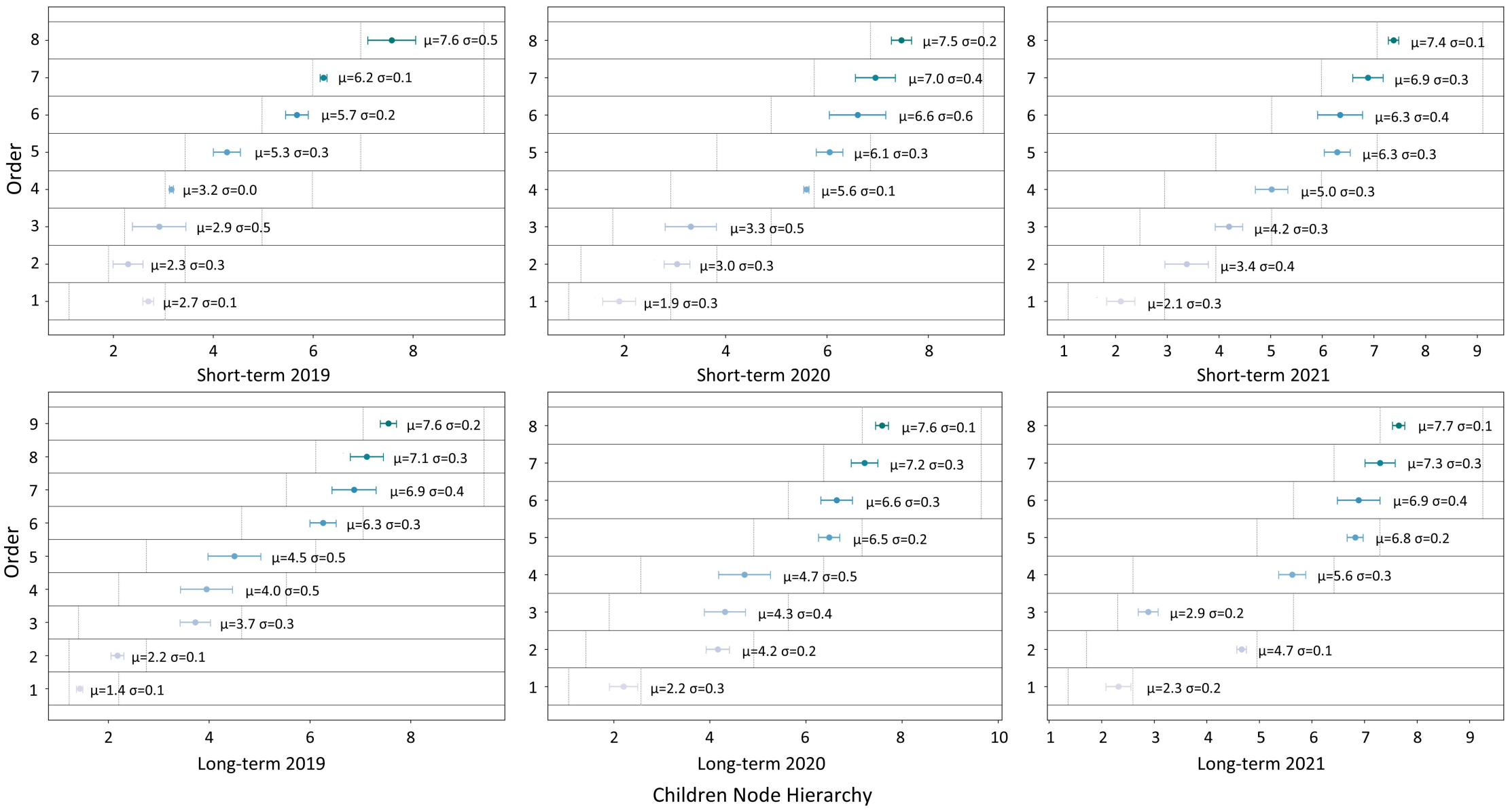

**Fig. S17. Distributions of hierarchy measures of the hinterland city that defines the range by city order for long-term and short-term datasets of China.** Each subplot shows results from a dataset. For each order, the mean (μ) and standard deviation (σ) of the hierarchy measures are displayed. Due to the maximum limit across city orders, the vertical gray dashed lines represent the ideal boundaries of the child city hierarchy distribution.